\documentclass{osa-article}

\journal{oe}

\newcommand{\ignore}[1]{}
\usepackage{tikz}
\usetikzlibrary {arrows}
\usetikzlibrary{arrows.meta}
\usepackage[export]{adjustbox}
\usepackage{caption}
\usepackage{subcaption}
\usepackage{graphicx}
\usetikzlibrary{fit}
\usepackage{textgreek}
\usepackage[hang]{footmisc}
\usepackage{cleveref}
\usepackage{mathrsfs,amsmath}
\usepackage{pifont}
\usepackage{booktabs}
\usepackage{graphicx}
\usepackage[table]{xcolor}
\usepackage{longtable}
\usepackage{makecell}
\usepackage{array}
\usepackage{float}

\usepackage{todonotes}

\articletype{Research Article}

\begin{document}

\title{Advances in mid-infrared spectroscopy enabled by supercontinuum laser sources}

\author{Ivan Zorin,\authormark{1,*} Paul Gattinger,\authormark{1} Alexander Ebner,\authormark{1} and Markus Brandstetter\authormark{1} \ignore{ and Author Three\authormark{2,3} } }

\address{\authormark{1}Research Center for Non-Destructive Testing, Science Park 2,
Altenberger Str.69, 4040 Linz, Austria\\
}

\email{\authormark{*}ivan.zorin@recendt.at} 



\begin{abstract}
Supercontinuum sources are all-fiber pulsed laser-driven systems that provide high power spectral densities within ultra-broadband spectral ranges.
The tailored process of generating broadband, bright, and spectrally flat supercontinua\textemdash through a complex interplay of linear and non-linear processes\textemdash has been recently pushed further towards longer wavelengths and has evolved enough to enter the field of mid-infrared (mid-IR) spectroscopy.
In this work, we review the current state and perspectives of this technology that offers laser-like emission properties and instantaneous broadband spectral coverage comparable to thermal emitters. We aim to go beyond a literature review. Thus, we first discuss the basic principles of supercontinuum sources and then provide an experimental part focusing on the quantification and analysis of intrinsic emission properties such as typical power spectral densities, brightness levels, spectral stability, and beam quality (to the best of the authors' knowledge, the M\textsuperscript{2} factor for a mid-IR supercontinuum source is characterized for the first time). On this basis, we identify key competitive advantages of these alternative emitters for mid-IR spectroscopy over state-of-the-art technologies such as thermal sources or quantum cascade lasers.
The specific features of supercontinuum radiation open up prospects of improving well-established techniques in mid-IR spectroscopy and trigger developments of novel analytical methods and instrumentation. The review concludes with a structured summary of recent advances and applications in various routine mid-IR spectroscopy scenarios that have benefited from the use of supercontinuum sources.
\end{abstract}

\section{Introduction}
Mid-infrared (mid-IR) spectroscopy is a lab-standard non-destructive analytical technique that enables qualitative and quantitative chemical analysis of samples in all states of aggregation. 
The keystone of mid-IR spectroscopic methods is the study of the interaction of polychromatic mid-IR radiation with the matter under examination. The mid-IR spectral band spans the range from 4000~cm\textsuperscript{-1} to 400~cm\textsuperscript{-1} (2.5~\textmu m-~25~\textmu m).
Interpretation and analysis of measured spectra provide insight into the structure of molecules (e.g., atomic masses and bond strengths) and enable quantification of concentration\cite{chalmers_handbook_2001}.
Hence, mid-IR light sources are the workhorses that form the technological ground for mid-IR spectroscopic measurements. This work reviews the novel technology of mid-IR supercontinuum laser sources as an alternative to state-of-the-art light emitters in this specific field.

In most routine scenarios, mid-IR spectrometers exploit thermal emitters (silicon carbide elements\textemdash Globars) that are well-developed, robust, stable, and cost-effective. These classical sources have reached their emission capabilities during the rapid evolution of the instrumental basis in the last century, which was primarily aimed at detectors\cite{Historyofinfrareddetectors} and advancements of methods (e.g. establishment of Fourier-domain IR spectroscopy, FTIR\cite{griffiths_fourier_2007}). Further improvement of thermal sources is challenging as it requires a significant temperature increase. However, it is possible, as demonstrated by the technology of non-classical laser-driven plasma sources\cite{doi:10.1021/acsphotonics.7b01484}. 
Thermal sources emit quasi-black-body radiation with a power spectral density (PSD) described by Planck's law. Their spectral coverage is well suited for most mid-IR spectroscopic applications. However, they impose several fundamental limitations. Thermal emitters are spatially incoherent, omni-directional and provide relatively low output optical power levels. This cumulatively leads to deficient brightness (the term owes its usage to laser physics\cite{svelto2010principles}, also known as spectral radiance). 
The practical importance of brightness should not be neglected in optical metrology, and particularly in spectroscopic measurements. The brightness of a light source directly affects the spectral power incident on the sample unit surface area. For mid-IR spectroscopy it means that, according to the Beer's law, the light-matter interaction path lengths can be extended preventing total light attenuation\textemdash more molecules can be investigated. Thus, limits of detection can be lowered. On the other hand, in photothermal applications, the signal is directly proportional to the intensity of the incident light. Here, sufficient brightness is even more beneficial.

The invention and development of the laser provided an advanced high-brightness source devoid of the disadvantages of thermal emitters\cite{svelto2010principles,SHUKLA201540}. Lasers possess high spatial coherence (the emission mode area is small and formed by spatially correlated field patterns in contrast to extended thermal sources), directionality, high output optical powers and beam qualities. As a result, the brightness of laser sources is several orders of magnitude higher than that of thermal emitters\cite{svelto2010principles,doi:10.1119/1.2344276}. Since lasers are quasi-monochromatic, they had to be widely adapted for mid-IR spectroscopic applications. The first systems were developed to provide tunable emission (tailoring the emission wavelength by modifying system parameters or environmental conditions). Thereby, tunable small band-gap lead-salt diode lasers covering the entire mid-IR spectral range\cite{TACKE1995447,adams_leadsalt_2001} have been applied for spectroscopic measurements already in the mid-60s. These systems have verified the expectations on high-brightness sources in mid-IR spectroscopy\cite{brewer_advances_1974,WERLE1998197,https://doi.org/10.1029/2002JD002208}. However, they were not widespread due to the relatively low power levels (typically several hundred microwatts), inferior beam properties (high divergence, astigmatic emission), and strict operation requirements (cryogenic cooling is necessary to achieve population inversion)\cite{Tittel2003}. Diode lasers based on lead-salt heterostructures were soon followed by antimonide diode lasers\cite{JJJ2003,JOULLIE2003621,YIN20076,SCHILT20043431}, laser sources based on parametric frequency conversion (optical parametric oscillators, difference frequency generators)\cite{PETROV2012536,C5CP07052J}, and adapted solid-state lasers\cite{https://doi.org/10.1002/lpor.200810076}. These laser systems have resolved the limitations of the first designs. They offered\textemdash depending on the class of the emitters\textemdash particular advantages such as high output powers, room-temperature operation, and relative simplicity of the design. Conversion systems and solid-state lasers were capable of fundamental mode emission. In terms of spectral coverage, the range from 2.3~\textmu m to about 5~\textmu m became available for lasers at this time. Nevertheless, probably the major milestone towards advanced laser-based mid-IR spectroscopy has been made with the development of the band engineered semiconductor laser (based on intersubband transitions), namely the quantum cascade laser (QCL)\cite{faist_quantum_1994}.
In contrast to the above-mentioned systems, high brightness mid-IR QCLs have demonstrated particular application advantages due to their robustness, stable operation at room temperatures, compact dimensions and wide spectral tunability (several chips can be united in one emitter to expand the spectral coverage further). QCLs feature a narrow spectral linewidth and relatively high output average powers. The QCL technology virtually covers the entire mid-IR range. Through these particular strengths, QCL systems have gained widespread use, facilitating highly sensitive mid-IR spectroscopic measurements since fundamental and strong absorption bands can be probed, which were poorly accessible at the time for other laser systems. As a result, QCLs have become a standard tool in laser-based mid-IR absorption spectroscopy. They led to development and improvement of methods in various molecular fingerprinting fields such as process monitoring, gas tracing, and biomedical spectroscopy, laboratory scenarios etc.\textemdash as evidenced by recent extensive scientific reviews\cite{C7CS00403F,doi:10.1366/14-00001,C8AN01306C}. 
Nowadays, progress in mid-IR laser spectroscopy has not stopped and continues to impress. For example, in addition to increased sensitivity through high brightness sources, spectral resolution and accuracy can also be significantly enhanced by using frequency-comb lasers in dual-comb spectroscopy\cite{Coddington:16}. 
Beyond lasers, infrared synchrotron sources are worth mentioning, as they offer high-brightness levels in ultra-broad spectral ranges~\cite{Duncan:83}. These sources are superior to conventional thermal emitters in various mid-IR spectroscopic applications\cite{MILLER2006846,nasse_high-resolution_2011,MARCELLI20121390}. However, the practical use of synchrotrons is limited and they are unlikely to become standard laboratory equipment due to inflexibility and extreme operating costs.
In the following sections, we exhibit the capabilities of supercontinuum sources that have laser properties and high brightness beyond that of synchrotrons, for this reason supercontinuum lasers are called table-top synchrotrons in~\cite{PETERSEN2018182}. Our analysis and comparisons (e.g. of brightness levels) focus on the most established and gold-standard representatives: the state-of-the-art QCLs and conventional thermal sources.

The phenomenon of supercontinuum generation\textemdash the process underlying supercontinuum sources\textemdash was first observed and reported in 1970~\cite{PhysRevLett.24.584} (the term has been introduced in\cite{PhysRevA.21.1222}) in an experiment, which has been enabled by high peak power lasers. A continuous optical spectrum spanning the entire visible range has been produced via spectral broadening during the propagation of intense picosecond laser pulses (Q-switched mode-locked Nd:Glass laser, 530~nm wavelength, frequency-doubled in a potassium dihydrogen phosphate crystal) through bulk borosilicate glass. 
Hereinafter, driven primarily by interests of spectroscopy, supercontinua have been generated in various bulk dielectric media\cite{dubietis_ultrafast_2017,dubietis_experimental_2019,alfano_supercontinuum_2006chapter}. The processes behind spectral broadening were intensively studied, interpreted and tailored. The most significant milestone in the commercialization and establishment of supercontinuum sources as versatile and flexible metrological tools was passed with the transition to another generation platform, namely optical fibers\cite{alfano_supercontinuum_2006,RevModPhys.78.1135,Genty:07,dudley_taylor_2010,alfano_supercontinuum_2016}.
During the last decades, supercontinuum sources have been rapidly evolving. Currently, the focus is still on taking control of the generation process and stabilization of their emission\cite{doi:10.1063/5.0053436} (potentially switching to all-normal dispersion pumping regime\cite{PhysRevLett.90.113904,Ramsay_2014,Genier:19,Rampur:19,kwarkye_-amplifier_2020,saini_coherent_2020,Eslami:20,adamu_noise_2020,Xing:18,Zhang2:19,Nguyen:20,Yuan:20}). At the same time, numerous efforts are made to increase optical output power levels (especially in the mid-IR spectral range\cite{4811103,Swiderski:14,SWIDERSKI2014189,Liu:14,Li:20}), and to extend the spectral coverage primarily to the mid-IR~\cite{steinmeyer_entering_2014,OB133,Yu:15,Moller:15,10.1117/12.2209253,16um,Hudson:17,Martinez:s,Zhang:19,LEMIERE2021104397}, but also to the ultraviolet\cite{jiang_deep-ultraviolet_2015,Wang:18,Poudel:19,adamu_noise_2020} range.
An impressive example of the progress in this field is the recent rapid evolution of fluoride ZBLAN [ZrF\textsubscript{4}-BaF\textsubscript{2}-LaF\textsubscript{3}-AlF\textsubscript{3}-NaF] fiber based supercontinuum generators. Initially starting with a few milliwatts of output optical power~\cite{1546047,Xia:06} these sources have recently reached levels of tens of watts\cite{Yang:19}.
A closer look at the mid-IR spectral window reveals that various aspects of the multifaceted topic of supercontinuum generation are being studied by groups from different fields. Significant efforts are currently being made to expand the operational spectral window. For instance, the implementation of cascaded schemes emerged as a promising solution to achieve high spectral performance, particularly in terms of bandwidth and brightness\cite{Kubat-2:14,Petersen:16,Nguyen:18,Petersen:Ca19,https://doi.org/10.1002/lpor.202000011,Woyessa:21}. Besides, studies of supercontinuum generation processes and their controlled tailoring\cite{Genty:02,Dudley:02,Genty:04,RevModPhys.78.1135,Dudley:08,Dudley:09,dudley_taylor_2010,Heidt:11,Erkintalo:10,PhysRevLett.90.113904,Pricking:10,Poletti:11,wetzel_real-time_2012,Driben:12,Nguyen:13,Driben:13,dudley_instabilities_2014,klimczak_direct_2016,Petersen:16,Zhao:17,Strutynski:17,Chen:18,Saghaei:18,voropaev_generation_2019,Genier:19,Engelsholm:19,Genier:20,kwarkye_-amplifier_2020} are still fundamental research topics. In order to generate and handle the long-wavelength supercontinua, glass materials are being optimized. Material research has mainly focused on glasses with wide mid-IR transparency: tellurite\cite{Klimczak:19}, fluoride\cite{doi:10.1063/1.3254214,Liu:14,Yang:19,Yang:13,Salem:15,SWIDERSKI2014189,gauthier_mid-infrared_2018}), and chalcogenide glasses\cite{7076577,GATTASS2012345,Marandi:12,Granzow:13,Shabahang:14,Kedenburg:15,Robichaud:16,Kubat:16,Petersen:17,Wang:17,https://doi.org/10.1111/jace.14391,OB133,Kubat:14,Kubat-2:14,Cheng:16,SHIRYAEV2013225,wu_chalcogenide_2018,LEMIERE2021104397}. Moreover, already developed system architectures are being optimized and technically adapted\cite{AliRezvani:20}. As a result, several reports of ultra-broadband and bright supercontinuum generation in the mid-IR have recently been presented\cite{Martinez:s,OB133,LEMIERE2021104397,Huang:21}.
In contrast, this work reviews the current maturity state of the mid-IR fiber-based supercontinuum technology beyond these generation-related subjects\textemdash in the applied field of mid-IR spectroscopy\textemdash and discusses core concepts, principles, particular emission properties as well as prospects and potential application scenarios. 



\section{Mid-IR supercontinuum technology}
The modern mid-IR supercontinuum laser source is a prominent representative of fiber-based emitters\cite{Jackson:20}. Mid-IR supercontinua provide instantaneous ultra-broadband spectral coverage (more than an octave). The supercontinuum generation process is rooted in the complex meshing and co-action of linear and nonlinear processes occurring during the propagation of intense pulses in the optical fiber. Depending on the pump scheme, material parameters, fiber geometry, dispersion regime, and input pulse duration, the ensemble of phenomena and mechanisms responsible for spectral broadening can significantly vary, certain processes can dominate or be suppressed by others. The major nonlinear contributors to the process of supercontinuum generation are\cite{ALEXANDER2012349}: 
stimulated Raman scattering, self-phase modulation, four-wave mixing, modulation instability, cross-phase modulation, soliton dynamics (soliton fission and soliton self-frequency shift) and dispersive wave generation (extensive overviews on this topic can be found e.g. in\cite{Genty:07,agrawal2007nonlinear,alfano_supercontinuum_2006,dudley_taylor_2010}).

\begin{figure}[hbt]
\begin{tikzpicture}
\centering
 \node[anchor=south west,inner sep=0] (image) at (0,0,0) {\includegraphics[width=1\linewidth]{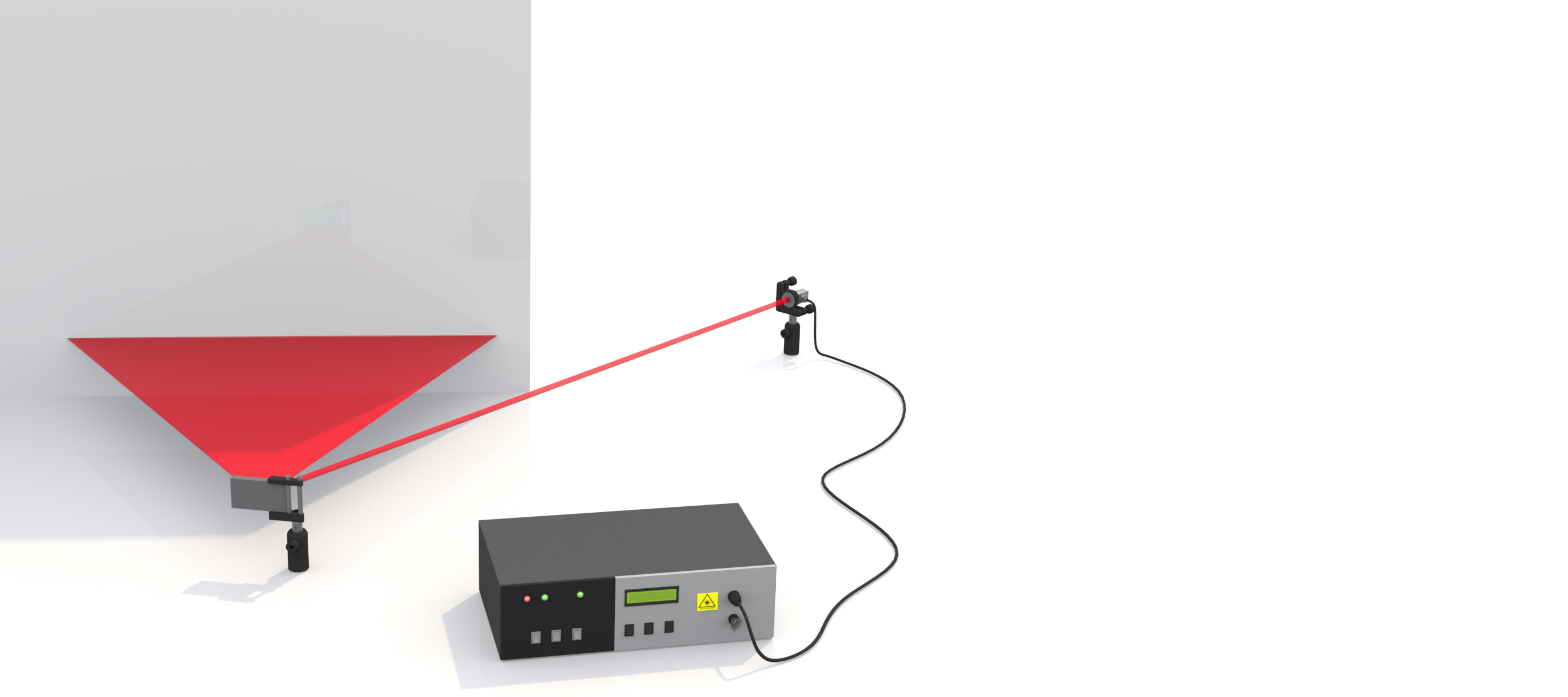}};
 \begin{scope}[x={(image.south east)},y={(image.north west)}]
  \node[] at (0.168,0.76) {\resizebox{4.1cm}{2.45cm}{\includegraphics[scale=0.15]{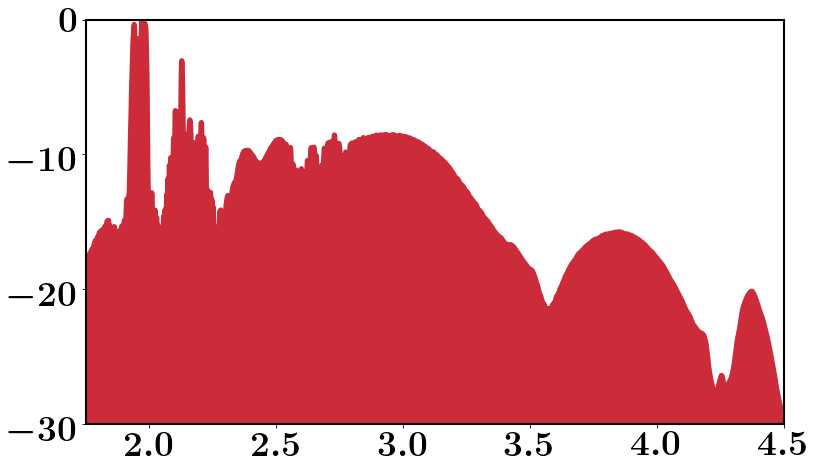}}};
  \node[] at (0.8,0.3) {\includegraphics[scale=1]{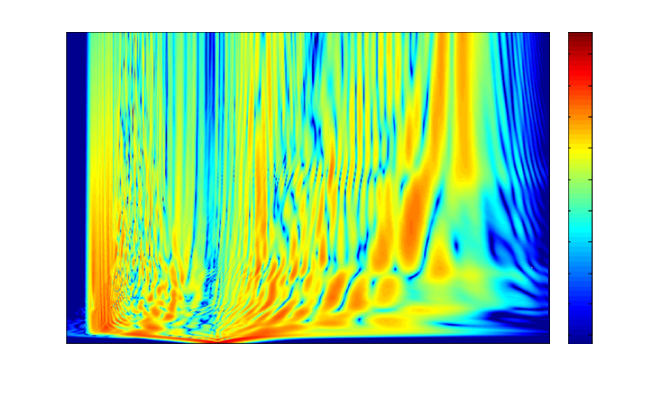}};
 \draw (0.38,0.345) node[]{\color{black}\scriptsize Supercontinuum source};
  \draw (0.38,0.3) node[]{\color{black}\scriptsize pump system};
   \draw (0.17,0.975) node[]{\color{black}\scriptsize Emission spectrum};
  \draw (0.78,0.57) node[]{\color{black}\scriptsize Spectral evolution within InF\textsubscript{3} fiber};
         \node at (0.7-0.022,0.84-0.0122) {\includegraphics[scale=0.4]{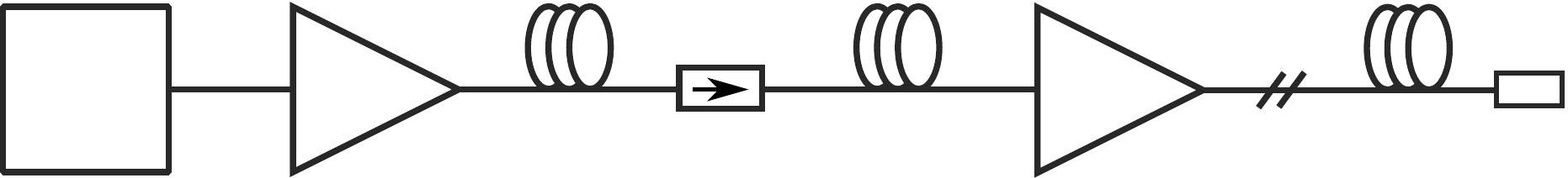}};
         \draw (0.44-0.022,0.95-0.0122) node[]{\color{black}\scriptsize Pump laser (fs)};
         \draw (0.445-0.022,0.84-0.0122) node[]{\color{black}\tiny 1550 nm};
         \draw (0.55-0.022,0.98-0.0122) node[]{\color{black}\scriptsize Er-doped};
         \draw (0.55-0.022,0.945-0.0122) node[]{\color{black}\scriptsize amplifier};
         \draw (0.62-0.022,0.77-0.0122) node[]{\color{black}\scriptsize Nonlinear};
         \draw (0.62-0.022,0.735-0.0122) node[]{\color{black}\scriptsize fiber};
         \draw (0.678-0.022,0.93-0.0122) node[]{\color{black}\scriptsize Isolator};
         \draw (0.74-0.022,0.77-0.0122) node[]{\color{black}\scriptsize Dispersion};
         \draw (0.74-0.022,0.735-0.0122) node[]{\color{black}\scriptsize compensating};
         \draw (0.74-0.022,0.7-0.0122) node[]{\color{black}\scriptsize fiber};
         \draw (0.82-0.022,0.98-0.0122) node[]{\color{black}\scriptsize Tm-doped};
         \draw (0.82-0.022,0.945-0.0122) node[]{\color{black}\scriptsize amplifier};
         \draw (0.872-0.022,0.77-0.0122) node[]{\color{black}\scriptsize Coupling};
         \draw (0.872-0.022,0.735-0.0122) node[]{\color{black}\scriptsize to fluoride};
         \draw (0.872-0.022,0.7-0.0122) node[]{\color{black}\scriptsize fiber};
         \draw (0.925-0.022,0.95-0.0122) node[]{\color{black}\scriptsize InF\textsubscript{3} fiber};
         \draw (0.97-0.022,0.77-0.0122) node[]{\color{black}\scriptsize Output};
         \draw (0.97-0.022,0.735-0.0122) node[]{\color{black}\scriptsize (collimator)};
        \draw[color={rgb,255:red,71; green,71; blue,71},thick,rounded corners=0.8pt] (0.37-0.022,0.67-0.0122) rectangle (1,1.0);
        \draw (0.418,1.04) node[fill={rgb,255:red,71; green,71; blue,71},rectangle]{\color{white}\scriptsize\textbf{Source diagram}};
  \draw (0.112,0.29) node[]{\color{black}\scriptsize Grating};
 \draw (0.112,0.25) node[]{\color{black}\tiny for illustrative};
  \draw (0.112,0.22) node[]{\color{black}\tiny purposes};
   \draw (0.175,0.54) node[]{\color{black}\tiny Wavelength (\textmu m)};
         \draw (0.01,0.75) node[rotate=90]{\color{black}\tiny PSD (dBm/nm)};
    \draw (0.5,0.64-0.0125) node[]{\color{black}\scriptsize Collimator};
    \draw (0.592,0.31) node[rotate=90]{\color{black}\scriptsize Propagation distance (cm)};
    \draw (0.5275,0.165) node[rotate=40]{\color{black}\scriptsize Nonlinear};
    \draw (0.535,0.375) node[rotate=40]{\color{black}\scriptsize InF\textsubscript{3} fiber};
    \draw (0.6182,0.535) node[]{\color{black}\scriptsize 200};
    \draw (0.6182,0.535-0.11) node[]{\color{black}\scriptsize 150};
    \draw (0.6182,0.535-0.22) node[]{\color{black}\scriptsize 100};
    \draw (0.6182,0.535-0.33) node[]{\color{black}\scriptsize 50};
    \draw (0.6182,0.535-0.44) node[]{\color{black}\scriptsize 0};
    \draw (0.636,0.07) node[]{\color{black}\scriptsize 1};
    \draw (0.636+0.0435,0.07) node[]{\color{black}\scriptsize 1.5};
    \draw (0.636+0.0435*2,0.07) node[]{\color{black}\scriptsize 2};
    \draw (0.636+0.0435*3,0.07) node[]{\color{black}\scriptsize 2.5};
    \draw (0.636+0.0435*4,0.07) node[]{\color{black}\scriptsize 3};
    \draw (0.636+0.0435*5,0.07) node[]{\color{black}\scriptsize 3.5};
    \draw (0.636+0.0435*6,0.07) node[]{\color{black}\scriptsize 4};
    \draw (0.636+0.0435*7,0.07) node[]{\color{black}\scriptsize 4.5};
    \draw (0.8,0.02) node[]{\color{black}\scriptsize Wavelength (\textmu m)};
    \draw (0.96,0.56) node[]{\color{black}\tiny (dBm/nm)};
    \draw (0.96,0.59) node[]{\color{black}\tiny PSD};

    \draw (0.979,0.505) node[]{\color{black}\tiny 5};
    \draw (0.979,0.505-0.0448) node[]{\color{black}\tiny 0};
    \draw (0.979,0.505-0.0448*2) node[]{\color{black}\tiny -5};
    \draw (0.979,0.505-0.0448*3) node[]{\color{black}\tiny -10};
    \draw (0.979,0.505-0.0448*4) node[]{\color{black}\tiny -15};
    \draw (0.979,0.505-0.0448*5) node[]{\color{black}\tiny -20};
    \draw (0.979,0.505-0.0448*6) node[]{\color{black}\tiny -25};
    \draw (0.979,0.505-0.0448*7) node[]{\color{black}\tiny -30};
    \draw (0.979,0.505-0.0448*8) node[]{\color{black}\tiny -35};
    \draw (0.979,0.505-0.0448*9) node[]{\color{black}\tiny -40};
 \end{scope}
\end{tikzpicture}
\caption{Basic scheme and operating principles of a mid-IR supercontinuum source exemplified for a InF\textsubscript{3} fiber based system: the depicted emission spectrum corresponds to the commercial supercontinuum generator (Thorlabs, SC4500, fiber length of 50~cm, 50 MHz repetition rate, 300~mW average output power); the simulated spectral evolution of the pump pulse over the 200~cm length InF\textsubscript{3} fiber\textemdash serves to illustrate the mechanism of generation; the source architecture and spectral evolution are adapted from~\cite{Salem:15} with permission from OSA.}
\label{fig:principle}
\end{figure}
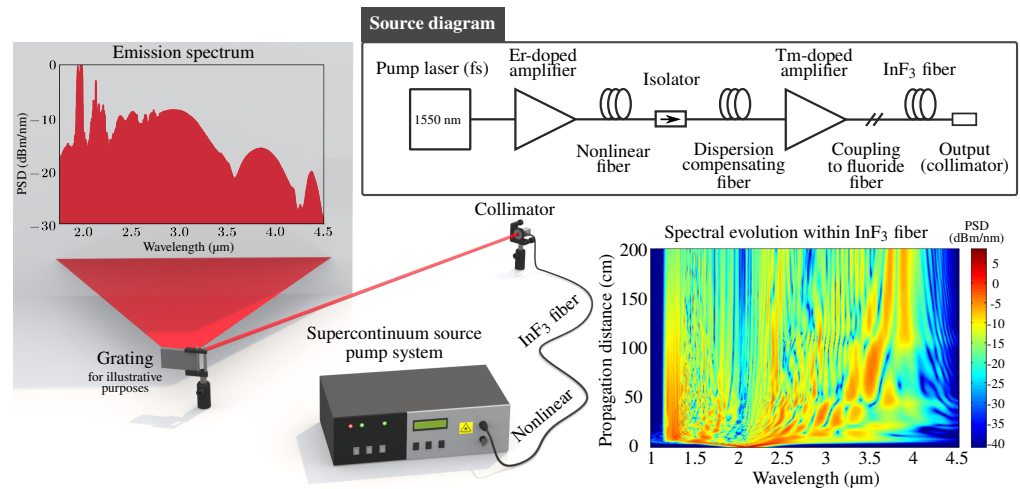

Despite the sophisticated fundamental physics behind supercontinuum generation, the practical realization for the mid-IR supercontinuum generation is relatively simple. Figure~\ref{fig:principle} illustrates this point and depicts the conceptual principles and system architecture of a commercial fluoride fiber based (InF\textsubscript{3}) supercontinuum generator. 
The system presented in Fig.~\ref{fig:principle} was developed, reported and described in~\cite{Salem:15}. The pump system of the supercontinuum source is based on a high peak power femtosecond mode-locked fiber laser. The laser radiates in the spectral range of the opto-communication wavelengths of 1550~nm, where optical technology is well-developed. The emitted laser pulses (50~MHz repetition rate) are amplified by an Er-doped fiber amplifier and launched into a nonlinear fiber that transfers the pulse energy to the 1.9~\textmu m spectral range, corresponding to the zero-dispersion wavelength of the designed fluoride fiber. The second amplification stage implies a boost of the optical power in the spectral range of around 2~\textmu m (to >0.5~W average power levels) using the following forward-direction Thulium-doped cladding-pumped fiber amplifier (793~nm pump diode). To compensate the anomalous group velocity dispersion of the Tm-doped and the delivery fibers, a dispersion compensating fiber is pre-employed finalizing the pump system dealing with the time-frequency adaptation of optical pulses for supercontinuum generation. Thereby, the shifted and spectrally pre-broadened pulses consisting of trains of solitons are being coupled into the 50~cm long InF\textsubscript{3} fiber, where substantial broadening occurs. The generated supercontinuum radiation spanning the spectral range from 1.25~\textmu m to 4.6~\textmu m is finally collimated by the output off-axis parabolic mirror.

The system presented in Fig.~\ref{fig:principle} exemplifies the spectacular phenomena of supercontinuum generation\textemdash how a relatively narrow, high-power near-IR laser line is being converted to an ultra-broadband and bright near- and mid-IR output.
Although the schemes, fiber types and design (for instance, small variations in core diameter can lead to significant changes in the emission spectrum due to variations in the dispersion profile\cite{Petersen:16}), pump parameters (duration, peak power, wavelength relative to the zero-dispersion point, polarization), number and realization of amplification stages can vary, the basic principle for modern supercontinuum generation that involves the pump laser and nonlinear fiber is preserved. In the same manner, typically high output powers (average power levels typically from 100~mW to watts), high beam qualities (typically M\textsuperscript{2}$\leq$1.1) and pulsed nature are also being maintained. 

From the perspective of mid-IR spectroscopic instrumentation and applications, high peak powers provided by mid-IR supercontinuum sources (a typical pulse duration is in the ps and sub-ns regimes) can significantly enhance detection in pulsed measurement modes, keeping average powers low and hence reducing the thermal load. Applying fast detection systems and suitable pulse integrators, such as lock-in amplifiers or boxcar gated averagers, the peak gain (the ratio of the peak to average powers, confined however by the response time of detector) can be exploited. It can be noted here that near-IR pure continuous-wave (CW) supercontinuum realizations have been reported but are not well-established and commercialized due to the inefficiency of the generation (efficiency of the nonlinear processes is suppressed by the meagre CW-power), poor broadening, requirements to extremely high average power pumping and km-long fibers resulting in subsequent challenges\cite{Nicholson2003,Choudhury:18}. 

\begin{figure}[ht]
\centering
 \begin{subfigure}[t]{.45\columnwidth}
  \includegraphics[height=5.05cm,trim=0 0 0 -3mm]{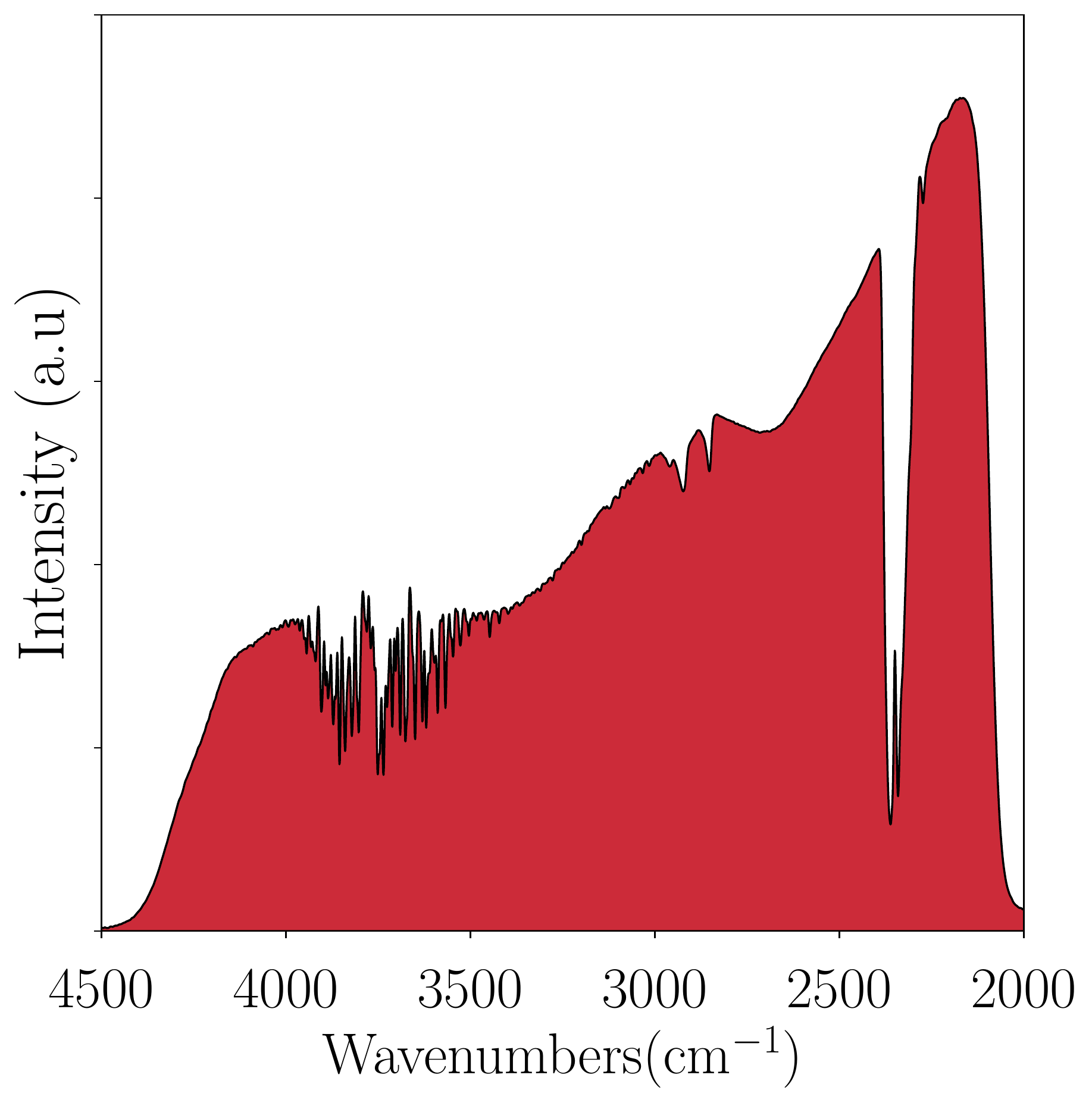}
  \caption{The mid -IR emission spectrum (2.4~\textmu m edge-pass filter is used) of Leukos InF\textsubscript{3} fiber based supercontinuum source (250 kHz repetition rate, 650~mW average output power)}
 \end{subfigure}
 \begin{subfigure}[t]{.45\columnwidth}
  \includegraphics[height=5cm]{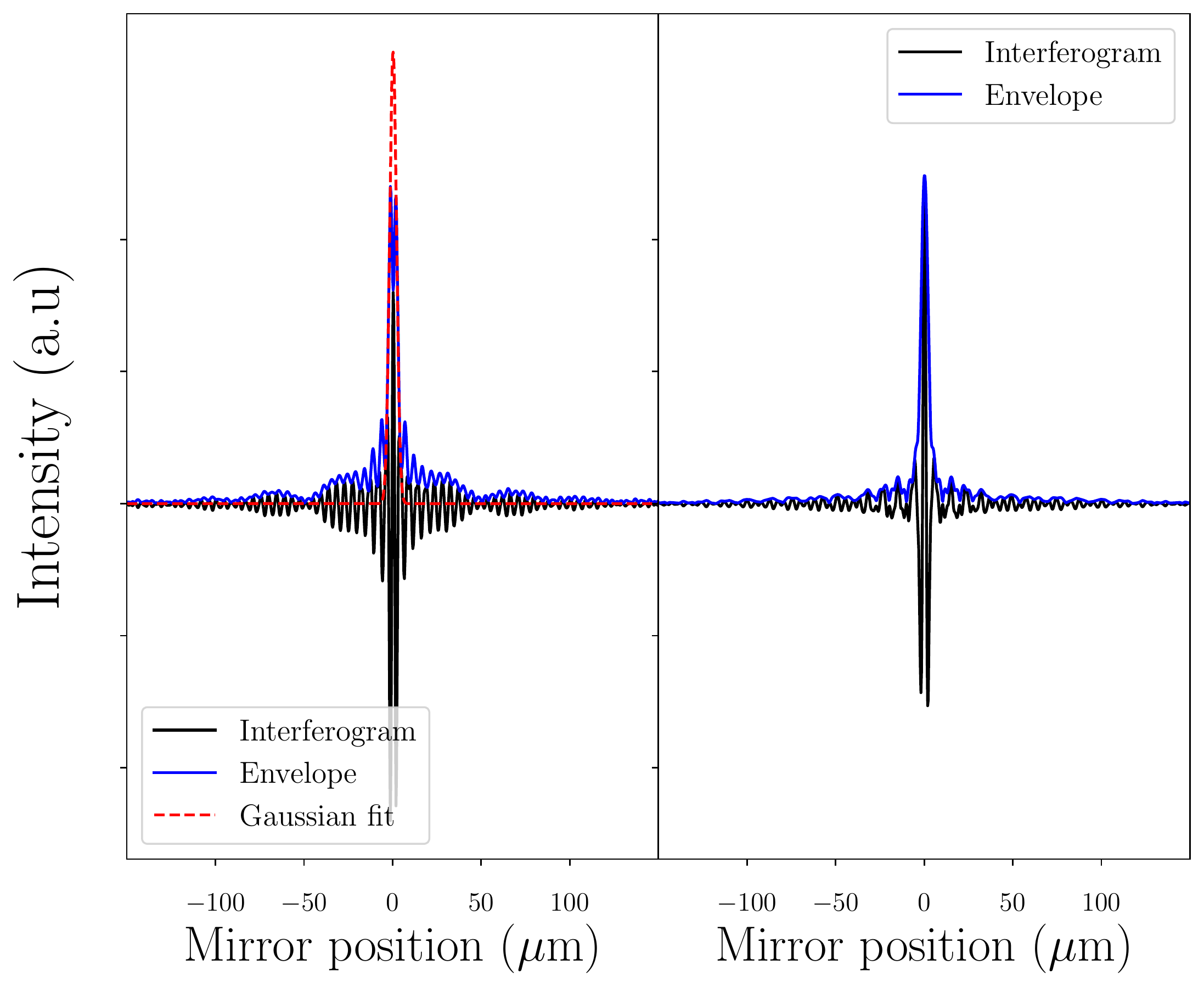}
  \caption{FTIR interferograms: (left) corresponding to the supercontinuum emission shown in (a) and (right) for the thermal emitter; analysis of autocorrelation functions (using the Hilbert transform) is used to access the actual coherence lengths}\label{fig:coherencelengths}
 \end{subfigure}
  \caption{Emission properties of a commercial supercontinuum emitter: (a) the mid-IR emission spectrum (atmospheric water vapour and CO\textsubscript{2} lead to the characteristic absorption features) and (b) temporal coherence properties (the interferogram for the conventional thermal emitter is shown for reference).} 
 \begin{tikzpicture}[overlay,thick]
 \draw [-latex,thick, color={black}] (2.06-0.8, 6.93+0.751) -- (2.06, 6.93+0.751);
 \draw [-latex,thick, color={black}] (2.1+0.8, 6.93+0.751) -- (2.1, 6.93+0.751);
 \draw (2.7, 7.1+0.751) node[]{\color{black}\scriptsize  $\approx$4.8~\textmu m};
 \draw [-latex,thick, color={black}] (4.755-0.5, 6.62+0.751) -- (4.755, 6.62+0.751);
 \draw [-latex,thick, color={black}] (4.8+0.5, 6.62+0.751) -- (4.8, 6.62+0.751);
  \draw (5.4, 6.8+0.751) node[]{\color{black}\scriptsize $\approx$4.8~\textmu m};
  \draw (2.1, 8.4+0.751) node[]{\color{black}\scriptsize Supercontinuum emission};
  \draw (4.76, 4.3+0.751) node[]{\color{black}\scriptsize Globar emission};
 \end{tikzpicture}
 \label{fig:coherence}
\end{figure}
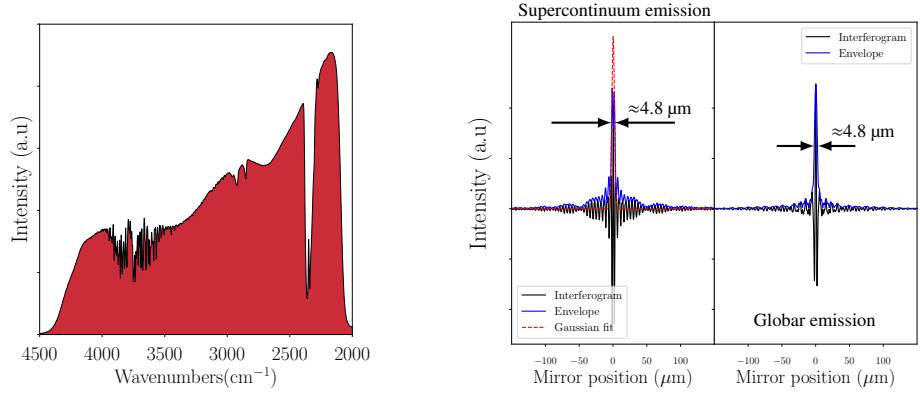

Apart from the features coming from the lasing nature, supercontinuum radiation has unique coherence properties. Supercontinua are generated within the guided mode of an optical fiber (often single mode)\textemdash thereby supercontinuum generators preserve high spatial coherence of the seed-pump laser resulting in high brightness and laser-like focusing properties\cite{Zeylikovich:05,Zeylikovich2006,genty_coherence_2016,alfano_coherence}. However, due to the extreme spectral broadening, the temporal coherence is being decoupled from the spatial one and undergoes changes during the generation process: supercontinuum light exhibits typically low temporal coherence (extremely short coherence lengths close to thermal light sources\ignore{; the degree of temporal coherence can vary for different implementations}).
Figure~\ref{fig:coherence} is intended to illustrate these points demonstrating the typical emission spectrum of a commercial system (Leukos InF\textsubscript{3} fiber based source) and a corresponding FTIR interferogram (i.e. field-autocorrelation). The Hilbert transform method was used to determine the coherence lengths shown in Fig.~\ref{fig:coherencelengths}. The obtained envelopes were used to extract the full width at half maximum values; a Gaussian fit was additionally applied to the interferogram envelope of the supercontinuum source due to the slightly asymmetric structure. The measurements were performed using a commercial FTIR spectrometer (Bruker Optics, Vertex 70) with default collection parameters (12 spectra averaged, 4~cm\textsuperscript{-1} resolution, 1~kHz mirror frequency). A boxcar integration (Zurich Instruments, UHFLI) was used to demodulate the signal.

\noindent Thus, these specific coherence and spectral peculiarities yield a unique emitter, which is of great interest in various mid-IR spectroscopy applications. Supercontinua eliminate temporal interference artifacts and maintain diffraction-limited performance, e.g. in hyperspectral imaging and microspectroscopy. For these reasons, these sources became of great interest also beyond mid-IR spectroscopy, for instance, in IR optical coherence tomography\cite{Cheung:15,Zorin:18,Israelsen:19,Zorin_OE:20,Zorin:21}.

In the following section, we address and quantify some supercontinuum emission properties that are particularly relevant for mid-IR spectroscopic applications. Furthermore, we sum up and feature application scenarios that have been or can be enhanced by novel mid-IR supercontinuum laser sources.

\section{Emission properties of mid-IR supercontinuum sources essential to spectroscopic applications}
\subsection{Brightness levels of mid-IR supercontinuum sources}
The definition of the brightness of light sources varies over different fields of science~\cite{SHUKLA201540}, therefore, in the following calculations and discussions, we employ the definition most familiar and specific to laser physics\cite{svelto2010principles}. 
The spectral brightness $\mathrm{B}_{\tilde\nu}$ of a given source is an emission property of the source that describes its spectral radiance (alternative naming for the brightness) and is defined as the average optical power $\mathrm{P}$ emitted into a certain direction per unit surface area $\partial S$ per unit solid angle $\partial\Omega$ per unit spectral line $\partial\tilde\nu$:
\begin{equation}
\label{eq:brightness}
\mathrm{B}_{\tilde\nu}(\tilde\nu)=\cfrac{\partial^3 \mathrm{P}}{\cos \theta~\partial S~\partial\Omega~\partial\tilde\nu},
\end{equation}
where $\theta$ is the polar angle between the normal to the surface and the vector that defines the direction of the emission. If $\mathrm{B}_{\tilde\nu}$ is independent of $\theta$, the source can be considered as isotropic and omni-directional\textemdash the most typical example here is thermal emitters with their spectral brightness given by Planck's law (used in the following calculations).

Equation~\eqref{eq:brightness} can be rewritten for laser beams to simplify calculations. Since for highly collimated laser beams the divergence angle $\theta$ is extremely small, we can assume $\cos\theta\approx 1$. Further, for a beam with the modal diameter $D$, the effective modal area is $\pi D^2/4$, and the emission solid angle can be defined\cite{svelto2010principles} as $\pi \theta^2$. Thereby, the spectral brightness of laser sources can be given as follows:
\begin{equation}
\label{eq:beam_brightness}
\mathrm{B}_{\tilde\nu}(\tilde\nu)=\cfrac{4~ \partial \mathrm{P}}{(\pi D \theta)^2 \partial\tilde\nu},
\end{equation}
where $\partial\mathrm{P}/\partial\tilde\nu$ is the power spectral density (PSD) of the source. Hence, brightness as a function of wavelength can be explicitly quantified, since PSDs, as well as beam sizes and divergences of various laser sources, are parameters that can be measured using standard techniques or provided by manufacturers. 

In order to stress the significance of spectral brightness for mid-IR spectroscopy, it should be considered in relation to optical instruments. The spectral brightness is conventionally expressed for wavelengths in units of W$\cdot$sr\textsuperscript{-1}$\cdot$cm\textsuperscript{-2}$\cdot$nm\textsuperscript{-1}. On the other side, the optical throughput of any mid-IR spectrometer (or etendue, a function of the system geometry and optical design, defined by the area of the entrance pupil and the solid angle formed by the collimating or the focusing optics) is usually given in units of sr$\cdot$cm\textsuperscript{2}\cite{saptari_fourier-transform_2003}. The product of multiplication of these characteristics of the light source and spectroscopic system is the spectral power that the certain system can transmit and exploit for spectroscopic measurements. Therefore, well-established mid-IR spectroscopic measurement methods can still be essentially improved by replacing a standard thermal emitter with ultra-bright supercontinuum laser emitters. For instance, maximum interaction lengths can be significantly extended, thus lowering limits of detection. In photothermal spectroscopy, signals can be significantly amplified.

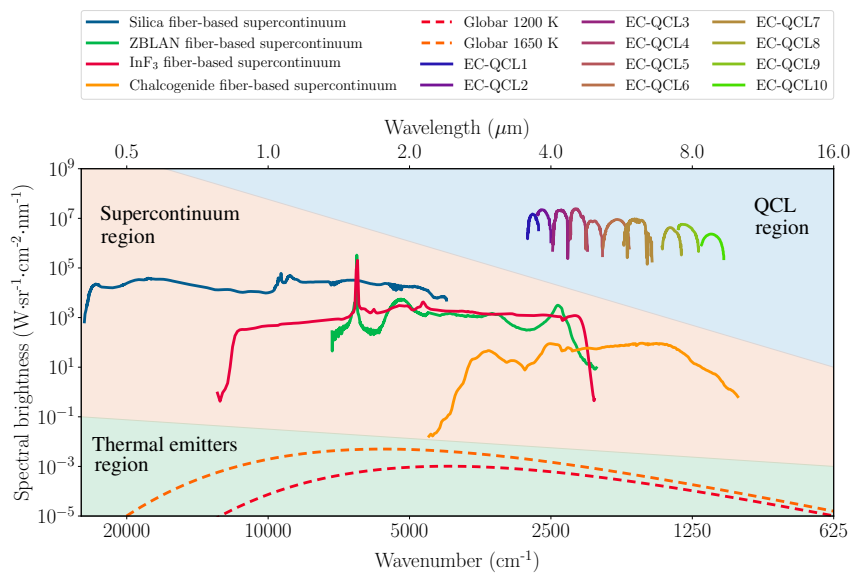
\begin{figure}[ht]
\centering
  \begin{tikzpicture}
\centering
 \begin{scope}[x={(image.south east)},y={(image.north west)}]
    \node at (0.5,0.5) { \includegraphics[width=0.85\columnwidth]{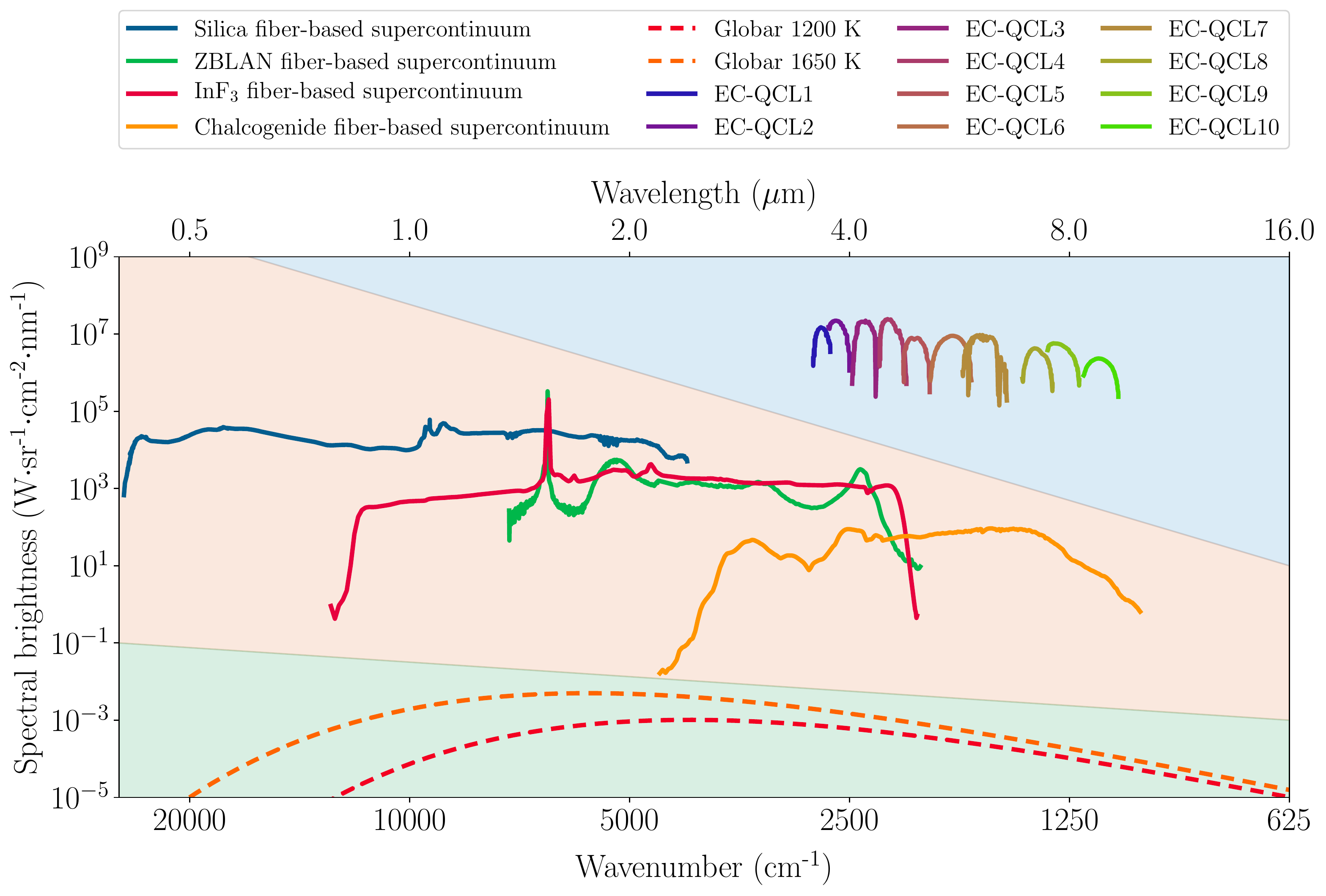}};
    \draw (0.843,0.69) node[]{\color{black}\footnotesize QCL};
    \draw (0.85,0.64) node[]{\color{black}\footnotesize region};
    \draw (0.24,0.67) node[align=center]{\color{black}\footnotesize Supercontinuum};
    \draw (0.1985,0.62) node[align=center]{\color{black}\footnotesize region};
    \draw (0.235,0.158) node[align=center]{\color{black}\footnotesize Thermal emitters};
    \draw (0.192,0.108) node[align=center]{\color{black}\footnotesize region};
 \end{scope}
\end{tikzpicture}
  \caption{Spectral brightness of most representative light sources in mid-IR spectroscopy (commercially available systems): spectral brightness levels of the Globars (thermal emitters) are determined using Plank's law; the represented supercontinuum sources are standard and not extreme power and bandwidth versions; EC-QCL - commercial external-cavity quantum cascade lasers (typical PSDs and beam parameters are used); the spectral brightness levels are calculated conventionally for average output power (the peak power advantage discussed below is not considered).}
 \label{fig:brightness_ir}
\end{figure}

Figure~\ref{fig:brightness_ir} provides a quantitative visualization illustrating typical brightness levels of supercontinuum sources operating in the IR range, as well as the brightness of gold-standard mid-IR spectroscopic sources: external-cavity QCLs and thermal emitters.

The brightness levels of the laser sources depicted in Fig.~\ref{fig:brightness_ir} were calculated according to Eq.~\eqref{eq:beam_brightness}. The PSDs, as well as the beam parameters of the silica-, InF\textsubscript{3} and chalcogenide-fiber based supercontinuum sources, were provided by Leukos. The emission parameters of the ZBLAN-fiber based supercontinuum source were provided by NKT Photonics. For the QCL emitters typical emission properties of commercial systems (DRS Daylight Solutions) were used; the CW operation mode was considered as it provides the narrowest spectral linewidth of 100~MHz.

Spectral brightness, as an explicit application-oriented metric to compare mid-IR emitters, clearly shows that ultra-broadband supercontinuum laser sources effectively fill the gap between thermal sources and QCLs. Thermal emitters are broadband but feature low brightness emission, while high-brightness QCL sources are relatively limited in spectral coverage, in particular, for the single laser ridge configuration. 
Thus, supercontinuum sources provide laser-like spectral brightness levels while maintaining an instantaneous ultra-broadband spectral coverage (close to thermal emitters in this sense) for each sub-ns pulse.

Considering the pulsed nature of the supercontinuum emission, another point has to be emphasized again in this context. The spectral brightness levels depicted in Fig.~\ref{fig:brightness_ir} are calculated for average powers that mostly heat the sample (typical PSD levels in \textmu W~-~mW ranges per nm-line) and not for the peak power levels. In the typical supercontinuum short-pulse emission regime, peak power levels can reach several watts within the same nanometer spectral band and thus peak brightness can be significantly higher. The conventional definition used above is completely valid for slow IR detectors. Nevertheless, fast semiconductor detectors (e.g. HgCdTe, InSb and InAsSb) can exhibit sub-ns rise time\cite{irdet}, thus, coupled to a suitable detection system (e.g. boxcar integrators) the peak power signal advantage can be gained. Hence, applying a proper detection scheme, pulsed supercontinuum sources can reach comparable brightness levels as for instance QCL lasers or even surpass them.


\subsection{Beam quality}

Emission features that significantly distinguish supercontinuum laser sources from standard emitters used in applied mid-IR spectroscopy\textemdash from both thermal and laser sources\textemdash are high beam quality and spatial stability. Thus, unlike, for example, QCL technology, mid-IR supercontinua generated in optical fibers have superior characteristics in terms of beam profile, divergence (beam qualities are severely limited for QCLs\cite{4431123,MROZIEWICZ2019161}), and beam symmetry (no astigmatism, common for QCL emission\cite{Tittel2003,4431123,MROZIEWICZ2019161}). Besides, supercontinuum sources provide an intrinsic mode-hop free operation.

The beam quality practically reflects the quality of the evolution of energy spatial distribution along the propagation direction. It not only directly affects the spectral brightness of the laser source through the divergence and modal area, but also determines the focusing performance. Thus, this emission feature removes the trade-off between spatial and spectral performance inherent for thermal emitters and becomes of particular interest for e.g. remote sensing and stand-off spectroscopic applications, chemical imaging and mapping, and microspectroscopy.

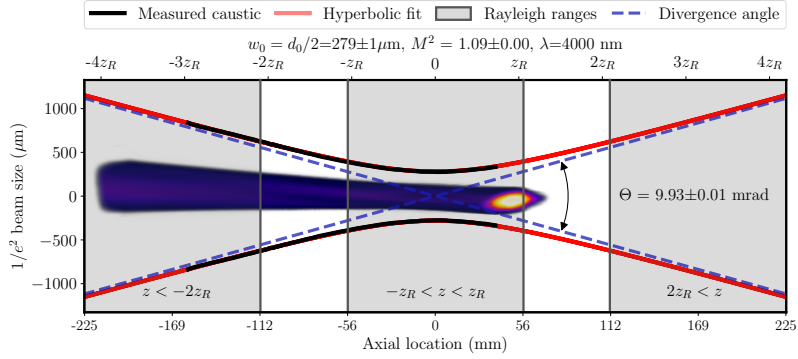
\begin{figure}[H]
\centering
\begin{tikzpicture}
\centering
 \begin{scope}[x={(image.south east)},y={(image.north west)}]
    \node at (0.42,0.47) {\includegraphics[scale=0.176]{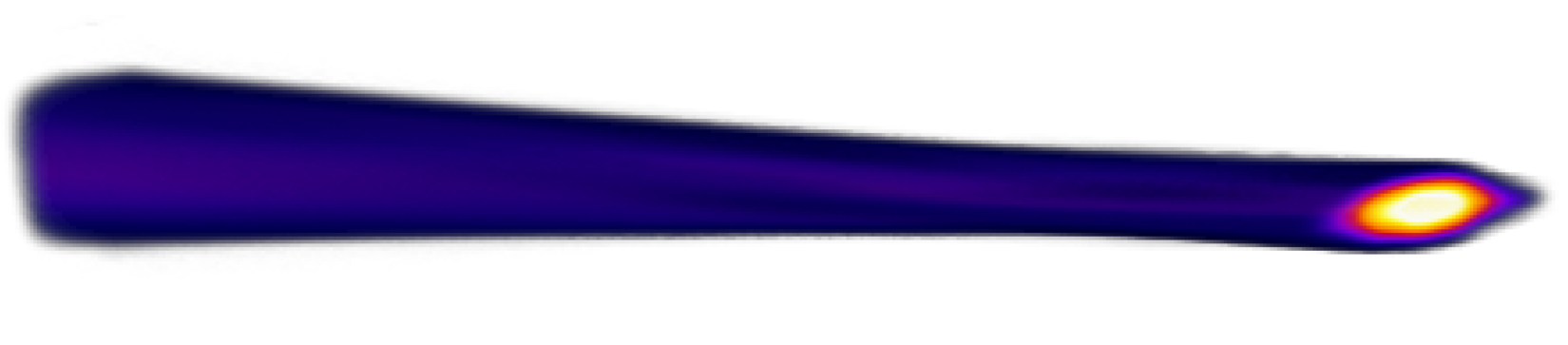}};
    \node at (0.5,0.5) {\includegraphics[width=0.8\columnwidth]{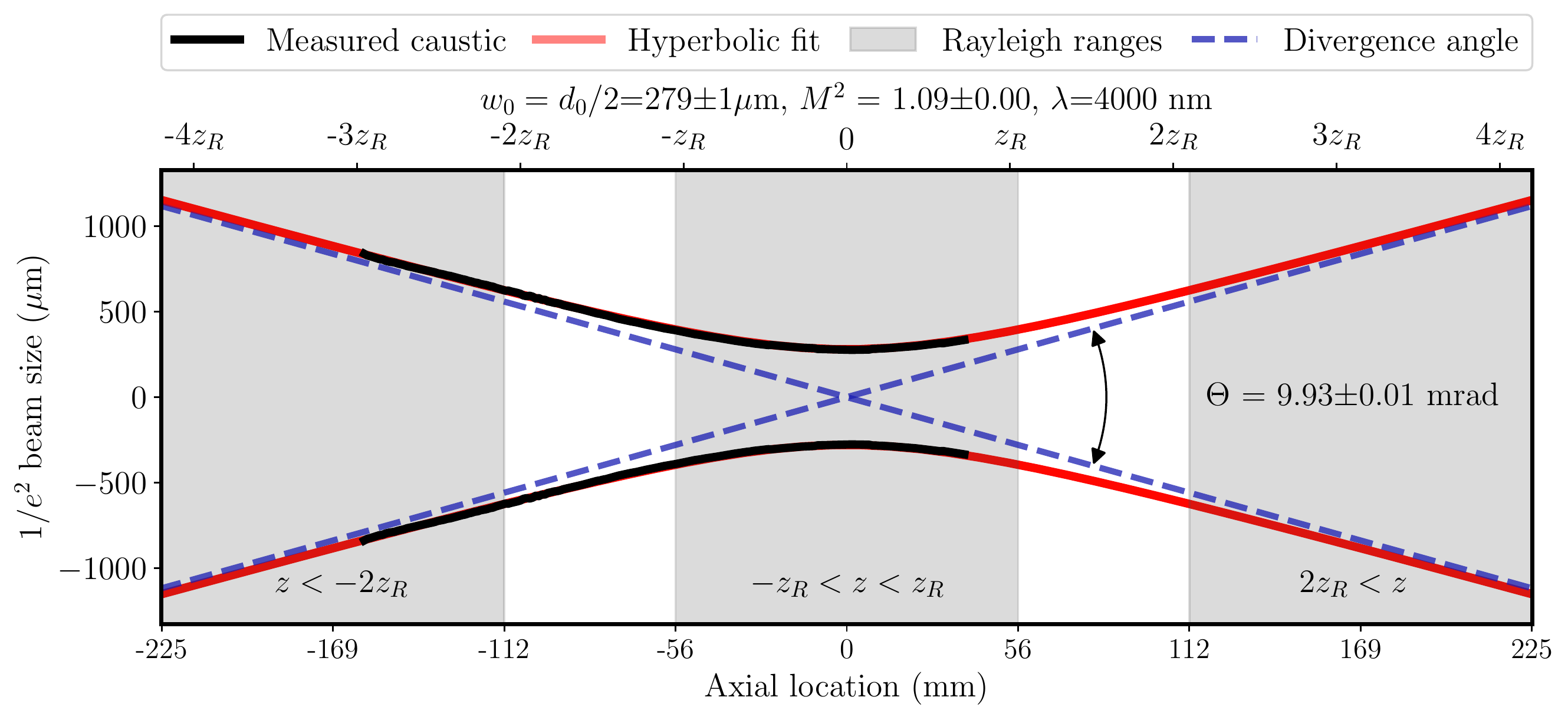}};
     \draw [-,line width=0.3mm, color={rgb:red,114;green,114;blue,114}] (0.358, 0.2) -- (0.358, 0.72);
     \draw [-,line width=0.3mm, color={rgb:red,114;green,114;blue,114}] (0.445, 0.2) -- (0.445, 0.72);
     \draw [-,line width=0.3mm, color={rgb:red,114;green,114;blue,114}] (0.62, 0.2) -- (0.62, 0.72);
     \draw [-,line width=0.3mm, color={rgb:red,114;green,114;blue,114}] (0.706, 0.2) -- (0.706, 0.72);
     \draw [-,color={rgb:red,114;green,114;blue,114},line width=0.3mm] (0.533,0.852) rectangle (0.5675,0.88);
 \end{scope}
\end{tikzpicture}
 \caption{Characterization of the M\textsuperscript{2} beam quality factor within the mid-IR sub-band (M\textsuperscript{2}~$\approx$~1.09, 500~nm band, 4~\textmu m center wavelength) for a commercial mid-IR ZBLAN-based supercontinuum source (NKT Photonics); the complete recorded 3-dimensional beam evolution (in pseudo colors) is shown in the background for reference (radial asymmetry is due to the spherical mirror that was used to avoid oversaturation); the analysis is performed for the meridional plane (the input beam profile is Gaussian).}
 \label{fig:M2_characterisation}
\end{figure}

In most applied cases, a practical parameter that characterizes and quantifies the laser beam quality sufficiently well is the M\textsuperscript{2} factor\cite{svelto2010principles}. It essentially indicates how strongly the actual beam differs from the theoretical diffraction-limited one (diffraction-limited Gaussian beam has an M\textsuperscript{2} factor of 1). The beam quality factor has an explicit practical meaning so that the resolution of, for example, any hyperspectral aberration-free microscope that employs the mapping approach can be estimated by multiplying the theoretical diffraction-limited resolution with the M\textsuperscript{2} factor of the exploited light source\cite{doi:https://doi.org/10.1002/0471213748.ch3}.

A procedure for determining the M\textsuperscript{2} factor is defined by the ISO Standard 11146\cite{ISO11146}. It involves the measurement of the beam caustic (for at least five beam locations within one Rayleigh distance [$z_R$] and five locations at distances more than two Rayleigh lengths from the waist) from which the M\textsuperscript{2} can be calculated analyzing the evolution of the beam radius accessed using the D4σ second moment of the intensity distribution method.

In this section, we provide the M\textsuperscript{2} characterization for a typical mid-IR ZBLAN-based supercontinuum source (NKT Photonics, SuperK Compact, 40~mW output power). 

In order to access the beam quality of the mid-IR supercontinuum emission, an all-mirror focusing optical arrangement (to eliminate chromatic aberration) was set up: a 750~mm focal length gold spherical mirror was used. The spectral range was limited using a suitable band-pass spectral filter (center wavelength is 4~\textmu m or 2500~cm\textsuperscript{-1}, 500~nm bandwidth, Thorlabs FB4000-500).
The beam profiles at different positions were recorded using a bolometer array (FLIR Boson, 640x480~px) fixed on a 20~cm scanning stage; the scanning range covers the necessary Rayleigh distances according to the ISO standard 11146. 

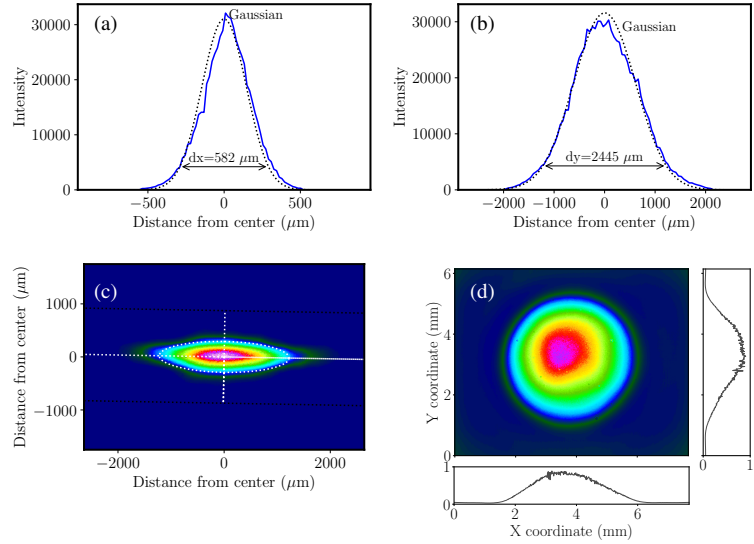
\begin{figure}[H]
\centering
\begin{tikzpicture}
\centering
 \begin{scope}[x={(image.south east)},y={(image.north west)}]
    \node at (0.707,0.15) {\includegraphics[scale=0.2]{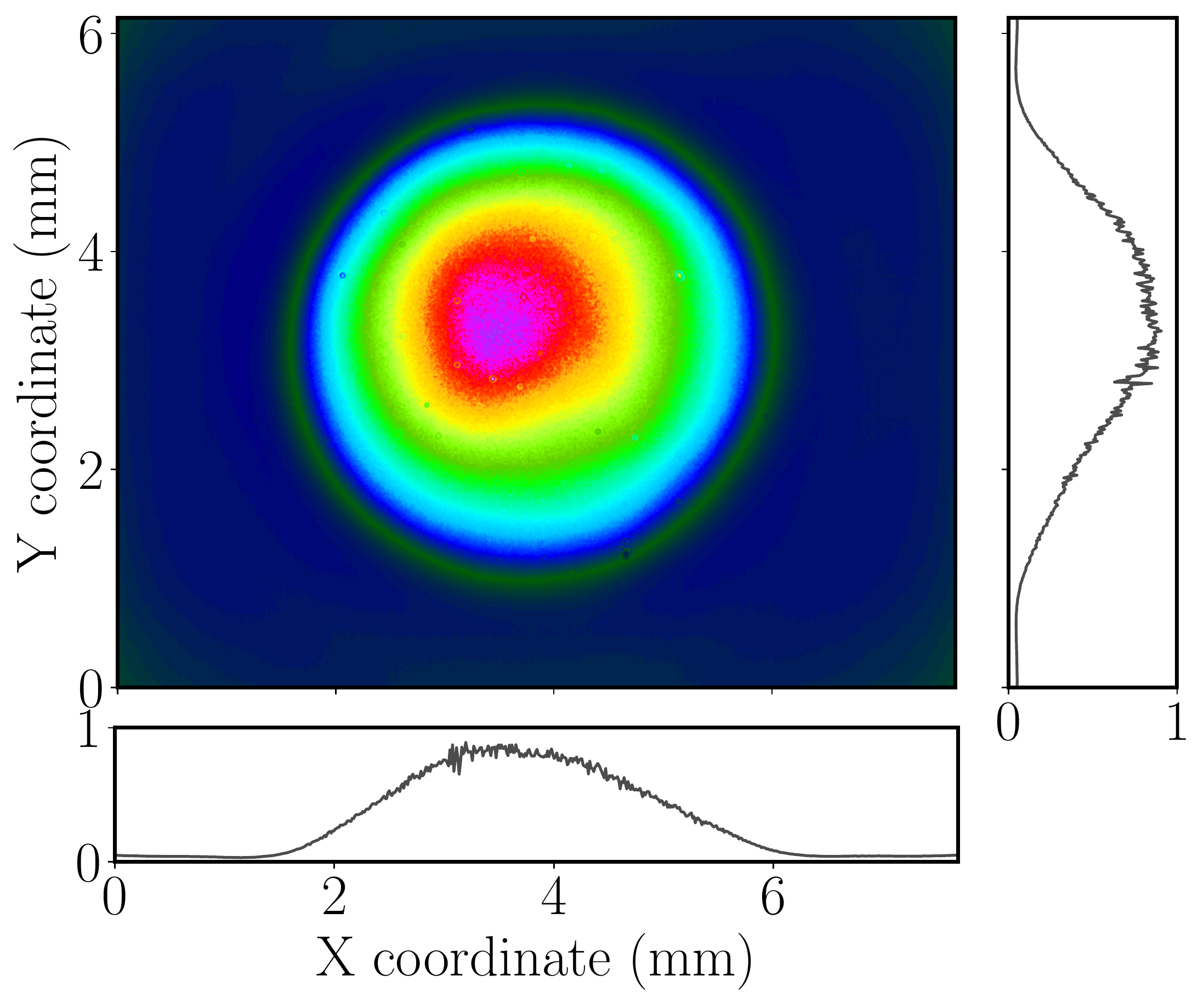}};
    \node at (0.5,0.5) {\includegraphics[width=0.75\columnwidth]{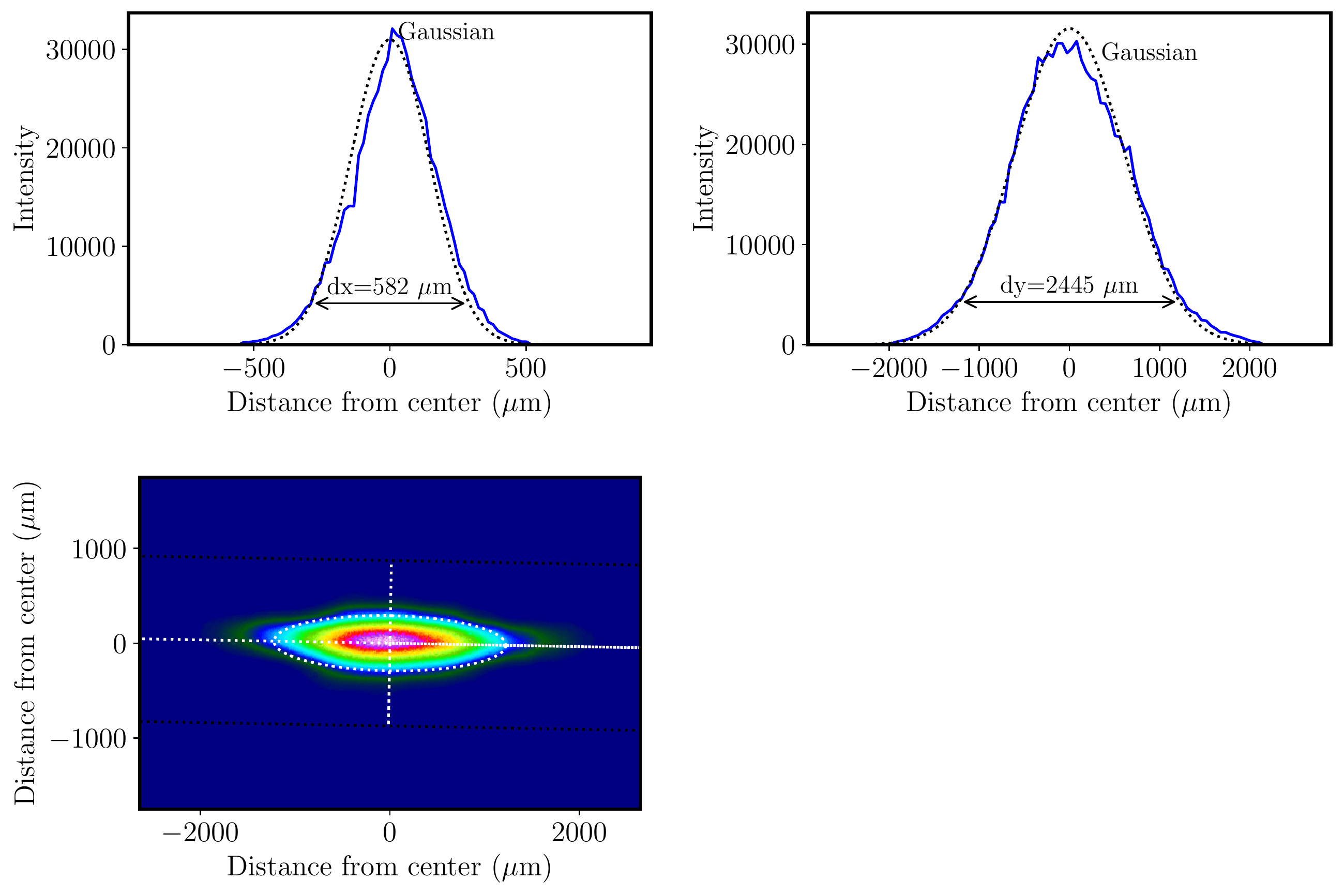}};
    \draw (0.225,1) node[]{\color{black}\footnotesize (a)};
    \draw (0.6,1) node[]{\color{black}\footnotesize (b)};
    \draw (0.225,0.4) node[]{\color{white}\footnotesize (c)};
    \draw (0.6,0.4) node[]{\color{white}\footnotesize (d)};
 \end{scope}
\end{tikzpicture}
 \caption{Profiles of the mid-IR supercontinuum beam obtained from the M\textsuperscript{2} characterization (4~\textmu m center wavelength, 500~nm bandwidth) at different positions: (a-c) close to the focus position (a,b are for major and minor axes of the on-purpose astigmated beam); (d) actual outgoing beam of the supercontinuum laser source measured directly after the collimator (normalized, the bolometer is not field corrected).}
 \label{fig:beamprofile}
\end{figure}

The supercontinuum beam caustic has been recorded with an axial step of 1~mm. Figure~\ref{fig:M2_characterisation} depicts the results of the measurements and characterization of the M\textsuperscript{2} (all the points within the [$-z_R<z<z_R$] and [$z<-2z_R$] ranges were used for calculation). The resulting M\textsuperscript{2} factor is 1.09, due to a large number of measurement points, the uncertainty is in the thousandths (0.0035). The postprocessing of the measurements was performed using the free Python library\cite{laserbeampy}.

Figure~\ref{fig:beamprofile} supplements the M\textsuperscript{2} characterization and displays a beam profile (with corresponding Gaussian fits for the semi-major and semi-minor axes) taken from the recorded caustics close to the focus position [see Fig.~\ref{fig:beamprofile}(a-c)]. A normalized polychromatic beam profile captured before the focusing system is shown in Fig.~\ref{fig:beamprofile}(c).

\subsection{Long-term stability}

\begin{figure}[ht]
\centering
 \begin{subfigure}[t]{.37\columnwidth}
 \includegraphics[width=\columnwidth]{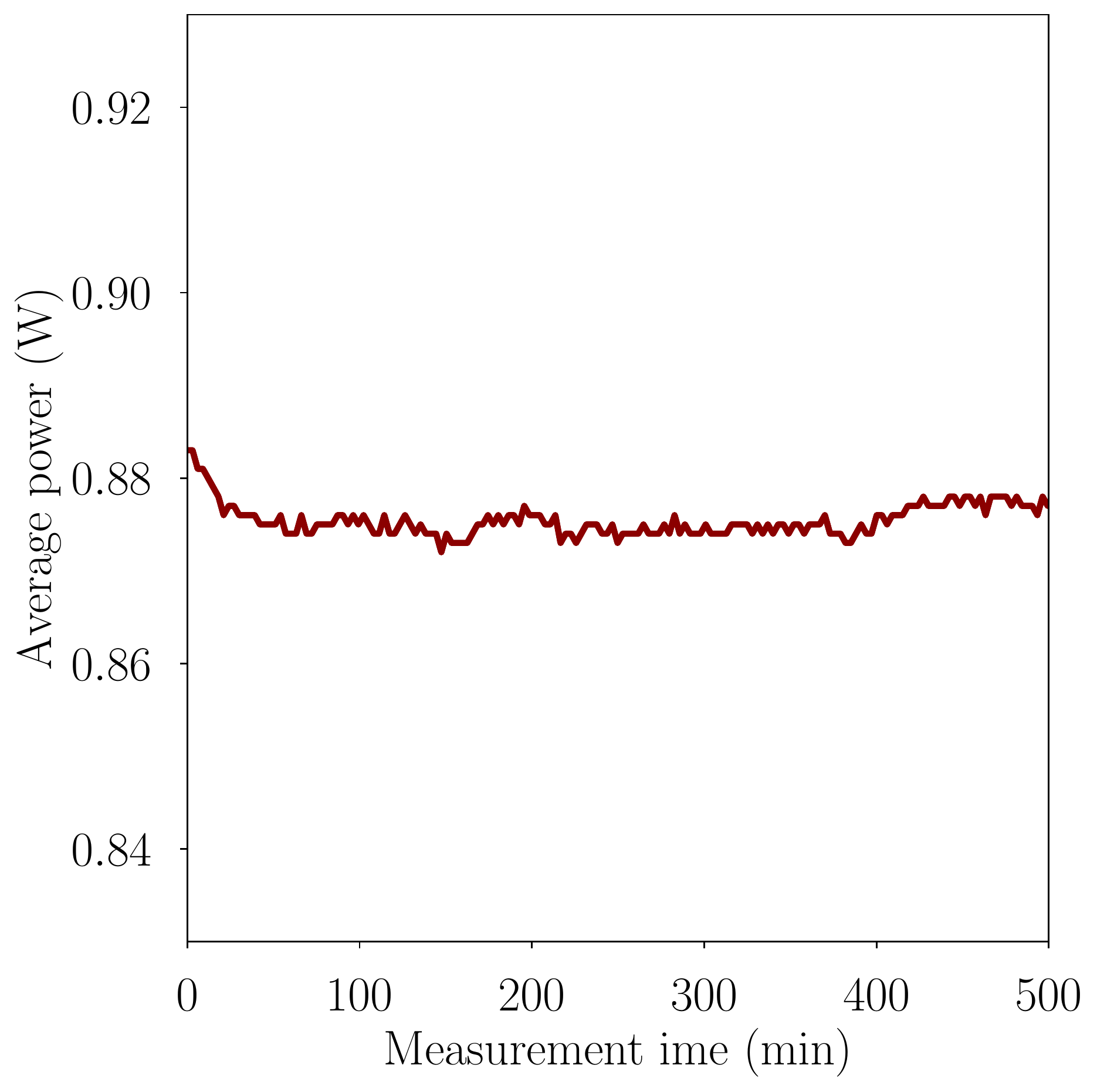} 
 \caption{Output power of a ZBLAN-based supercontinuum source (Novae, Coverage), recorded over 500 min (3~min sampling period); the mean value is 875~mW, the standard deviation is 1.7~mW}\label{fig:novae_power}
 \end{subfigure}
 \begin{subfigure}[t]{.37\columnwidth}
  \includegraphics[width=\columnwidth]{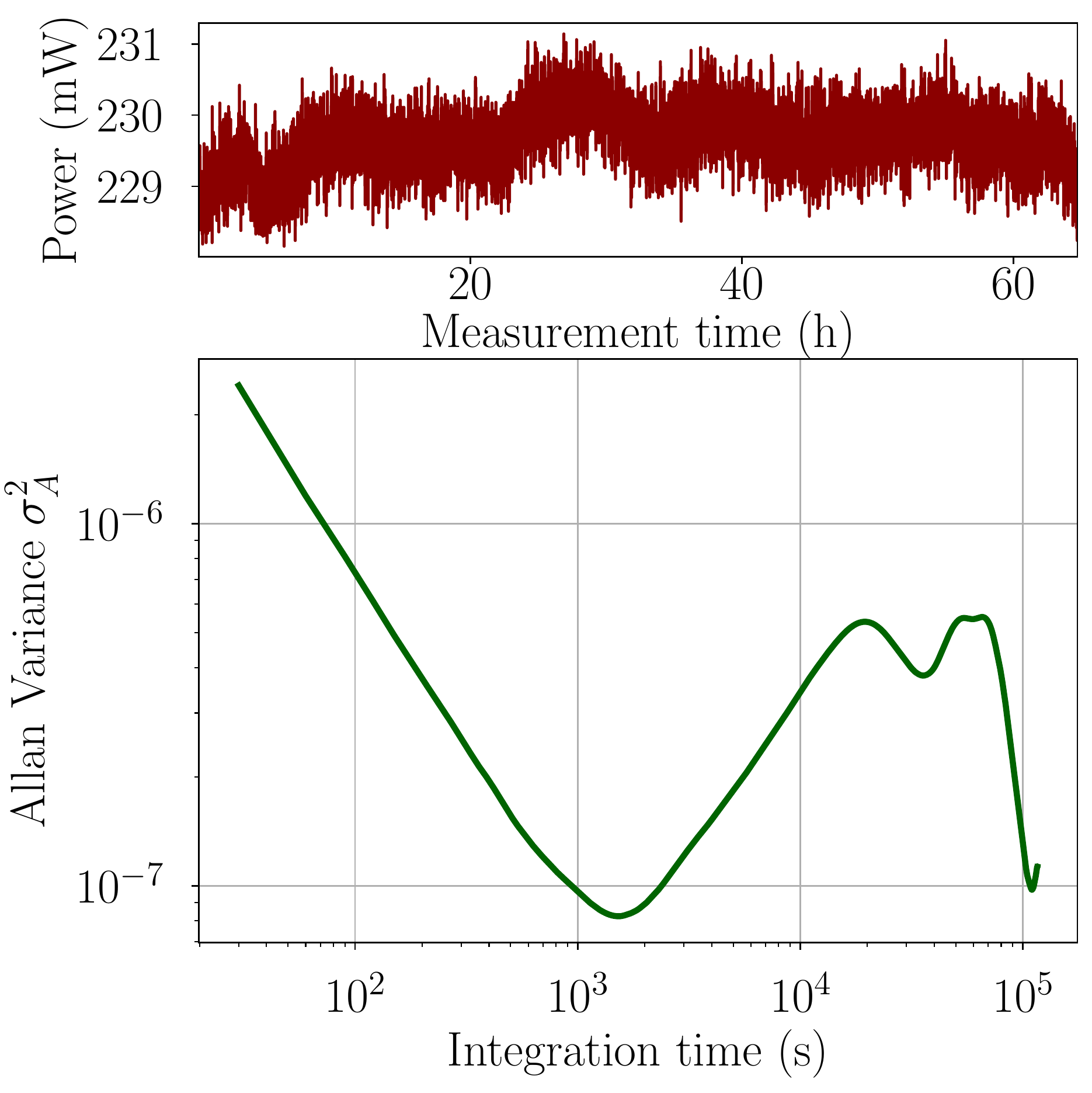}
  \caption{Allan variance plot (bottom) for the output intensity of a ZBLAN-based mid-IR supercontinuum source (NKT Photonics, SuperK mid-IR); calculated for the series of the recorded power levels (top);}\label{fig:allan}
 \end{subfigure}
  \caption{Long-term stability of the mid-IR supercontinuum sources characterized for commercial systems in the form of (a) the power stability (Novae supercontinuum emitter, 8~h) and (b) the Allan variance (NKT photonics, 65~h).}
 \begin{tikzpicture}[overlay,thick]
\draw (-2,8.7) node[]{\color{black}\scriptsize Stability: 99.8\% };
\draw (3,8.7) node[]{\color{black}\scriptsize Stability: 99.81\% };         
 \draw [dashed,thick, red] (1.2-0.05,6.8) to (2.8-0.13,4.45); 
 \draw [-,thick, red] (1.7,6.55) to (1.52,6.37); 
  \draw [-,thick, red] (2.875-0.14,4.51-0.14) to (2.875+0.14,4.51+0.14); 
  \draw [-,thick, red] (2.875-0.14,4.51+0.14) to (2.875+0.14,4.51-0.14); 
 \draw [-,thick, red] (2.875,4.6) to (2.875,5.3); 

\draw (2.3,6.65) node[]{\color{black}\scriptsize White noise };
 \draw (2.88,5.81) node[]{\color{black}\scriptsize Optimal };
  \draw (2.88,5.655) node[]{\color{black}\scriptsize integration };

 \draw (2.88,5.5) node[]{\color{black}\scriptsize time };
 \draw (3.65,5.2) node[rotate=57]{\color{black}\scriptsize Drifts };

 \end{tikzpicture}
 \label{fig:stabil}
\end{figure}
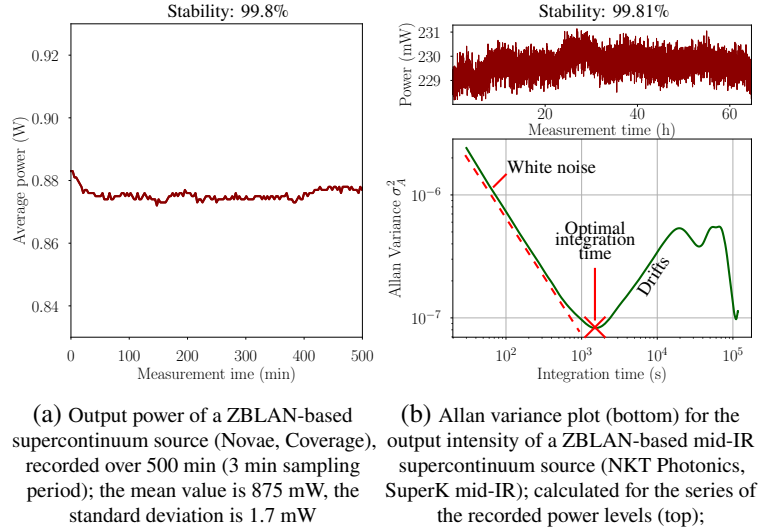

Besides the laser-like emission properties, modern mid-IR supercontinuum sources feature sufficient long-term intensity stability (see Fig.~\ref{fig:stabil}) that enables flexibility in measurements and integration without recalibration of spectroscopic instruments.

Long-term stability can be characterized in several ways. The most straightforward method is to assess the stability of the emission power over a reasonably long period of time. Figure~\ref{fig:novae_power} depicts the time-evolution of the output average power for a commercial mid-IR supercontinuum laser source (Novae, Coverage, 2.4~MHz repetition rate, spectral coverage from approx. 5200~cm\textsuperscript{-1} to 2380~cm\textsuperscript{-1}). The measurements were performed using a mid-IR compatible power meter (Coherent, LabMax TOP Power Meter, LM-10 HTD detector head) with a sample time step of 3~min. The relative output power fluctuations were 0.2\%, however, it should be noted that the first period of around 20~min caused stronger variations as the source showed a drift in the average output power during this warm-up time (excluding this time-span, the relative fluctuations are below 0.15\%).

\vspace{3pt}

In order to evaluate the long-term stability in terms of integration performances (i.e. to access the limits of averaging), distinguish potential noise sources and types of intensity fluctuations, an alternative measure can be employed. In spectroscopic measurements, the limit of detection is a function of the integration time coupled with the long-term stability of the instrument (i.e. mean values from an averaging process are being recorded, for instance, using e.g. pulse averaging or interferogram averaging). Thus, the concept of Allan variance\cite{AllanV,4404126} that is in essence the two-sample variance of the data cluster averages as a function of cluster size\textemdash first adopted for spectroscopy by Werle \cite{Werle1993,werle_accuracy_2011}\textemdash can be used.
In the following assessment, we employed the overlapping Allan variance estimator that, in contrast to the standard Allan variance algorithm used by Werle, exploits all possible combinations for the given data set by introducing overlapping clusters, and hence exhibits higher confidence\ignore{ (at the expense of higher computational costs)}\cite{807679,nistfreqs}. 
The following Allan variance estimator was used: 
\begin{equation}
\label{eq:oavar}
\sigma^2_A(k\cdot\tau_0 , N)=\cfrac{1}{2(N-2k+1)} \sum_{j=0}^{N-2k}  [A_{j+k}-A_j]^2,
\end{equation}
where $A_j$ is the average value of the \textit{j}\textsuperscript{th} cluster (also known as subgroups), $k$ is the cluster size (number of elements in the cluster), $\tau_0\cdot k$ is the observation time ($\tau=\tau_0\cdot k$, $\tau_0$ is the sampling period), $N$ is the total number of samples, the same notation as in\cite{Werle1993} is used here for simplicity.
The average value of the \textit{j}\textsuperscript{th} cluster is calculated then as:
\begin{equation}
\label{eq:avclust}
A_j=\cfrac{1}{k} \sum_{i=j}^{j+k}  x_i,
\end{equation}
where $x_i$ is the \textit{i}\textsuperscript{th} element of the data set.

Hence, the average output power sampled over a long period of time is chosen for the analysis here. It has to be noted, that this property was selected since the source fluctuations are isolated and the resulting Allan variances can be considered as a characteristic of the source. Thus, the Allan variance plot expresses the noise behavior of the supercontinuum source in a long run (regardless of the initial time moment) and defines average optimal integration times and noise reduction capabilities.

Figure~\ref{fig:allan} depicts the Allan plot for a ZBLAN-based mid-IR supercontinuum source (NKT Photonics, SuperK mid-IR, spectral coverage from approx. 6600~cm\textsuperscript{-1} to 2250~cm\textsuperscript{-1}, around 230~mW of mid-IR power, anomalous pump scheme). The variance was derived for a series of average power measurements, the sampling period was 30~sec, total evaluation time was 65~hours. 
The obtained Allan plot reveals that at low integration times, white noise dominates as the variance decreases proportionally to the averaging time. The origin of the observed white noise can be attributed to the characteristic property of the supercontinuum generation with pumping in the anomalous mode\cite{Dudley:02,klimczak_direct_2016}. 
The Allan plot also shows that the signal-to-noise ratio can be improved by more than one order of magnitude by averaging over the optimal time period. The observed long-term fluctuations start to contribute negatively beyond the optimal integration time of around 1500~sec reducing integration efficiency.
A second noise reduction can be observed for a cluster size larger than one day (non-expedient times for single spectroscopic measurements). These daily fluctuations can be correlated with e.g. temperatures changes.


\subsection{Spectral stability}

The presented long-term measurements reflect the time-stability of the average power of the supercontinuum sources and thus signal-to-noise enhancements achievable by increasing the measurement time. However, another metric that is of high interest for mid-IR spectroscopy is emitter-induced spectral fluctuations on time intervals on the scale of a standard measurement time. From this point of view, zero absorbance lines provide an insight into this property. They are calculated as Beer's absorbance for two sequentially measured spectra under the same conditions with no changes in the optical path and system. Thereby, they can be used to evaluate noises of different emitters as they show wavenumber-dependent instabilities. We performed these measurements with no sample inserted. Intensities were artificially scaled to comparable levels using neutral-density filters and different aperture sizes. Hence, the results do not represent the merit of the absolute signal-to-noise ratio available: e.g. by increasing path lengths for spectroscopic measurements, signals for thermal sources can be strongly or completely suppressed, while supercontinua can still efficiently penetrate. Quantitative assessments of spectral noise for various state-of-the-art laser-based mid-IR spectroscopic methods can be found in\cite{Abbas:21}.

\begin{figure}[H]
\centering
 \begin{subfigure}[t]{.45\columnwidth}
  \includegraphics[width=\columnwidth]{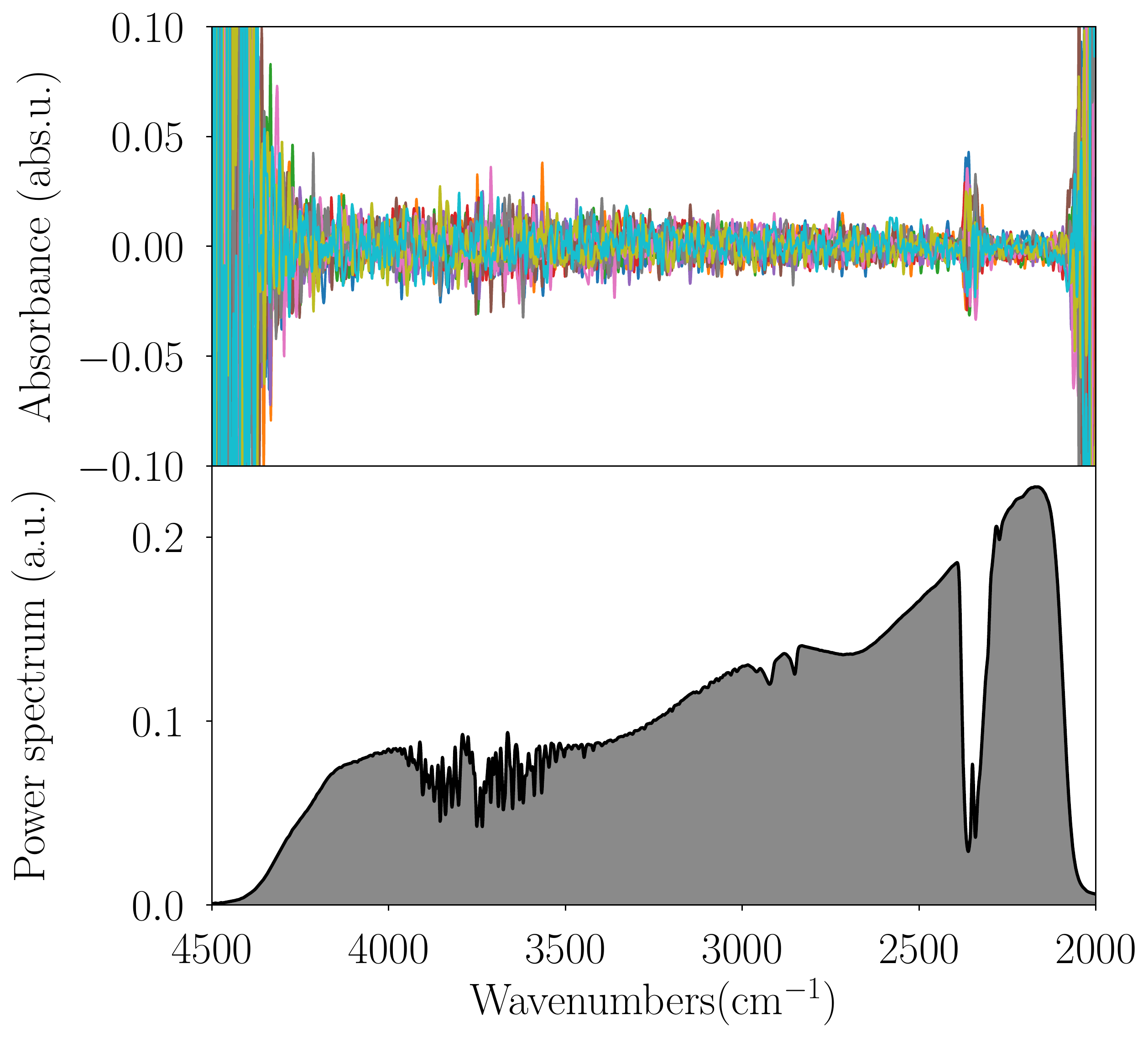}
  \caption{Leukos InF\textsubscript{3} fiber based supercontinuum source (600~mW, 250~KHz; 1\% ND filter used); \\ RMS=0.0065 (abs.u.)}
 \end{subfigure}
 \begin{subfigure}[t]{.45\columnwidth}
  \includegraphics[width=\columnwidth]{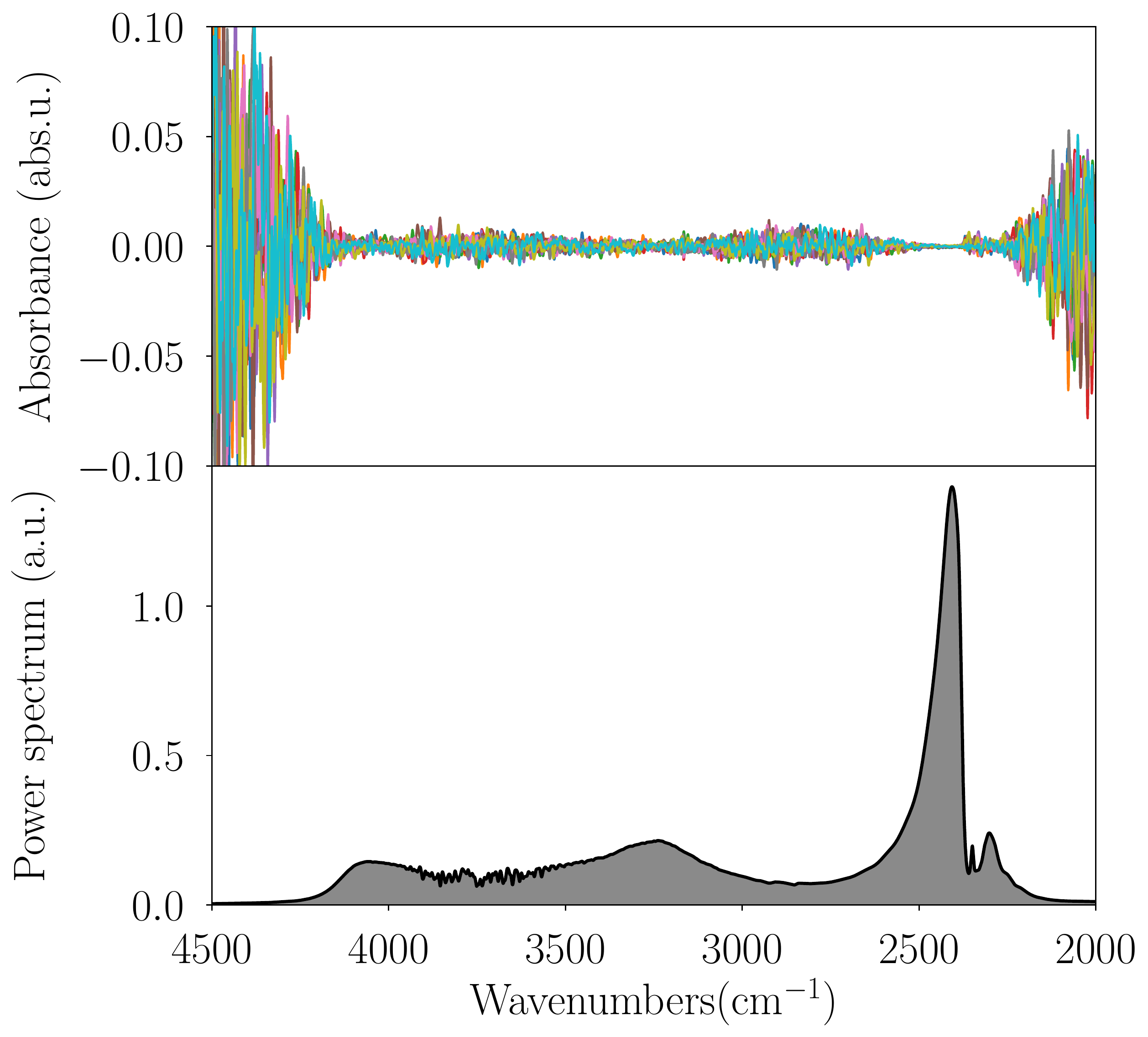}
  \caption{NKT photonics ZBLAN fiber based supercontinuum source (500~mW, 2.5~MHz); RMS=0.0025 (abs.u.)}\label{fig:nkt100lines}
 \end{subfigure}
 
  \begin{subfigure}[t]{.45\columnwidth}
  \includegraphics[width=\columnwidth]{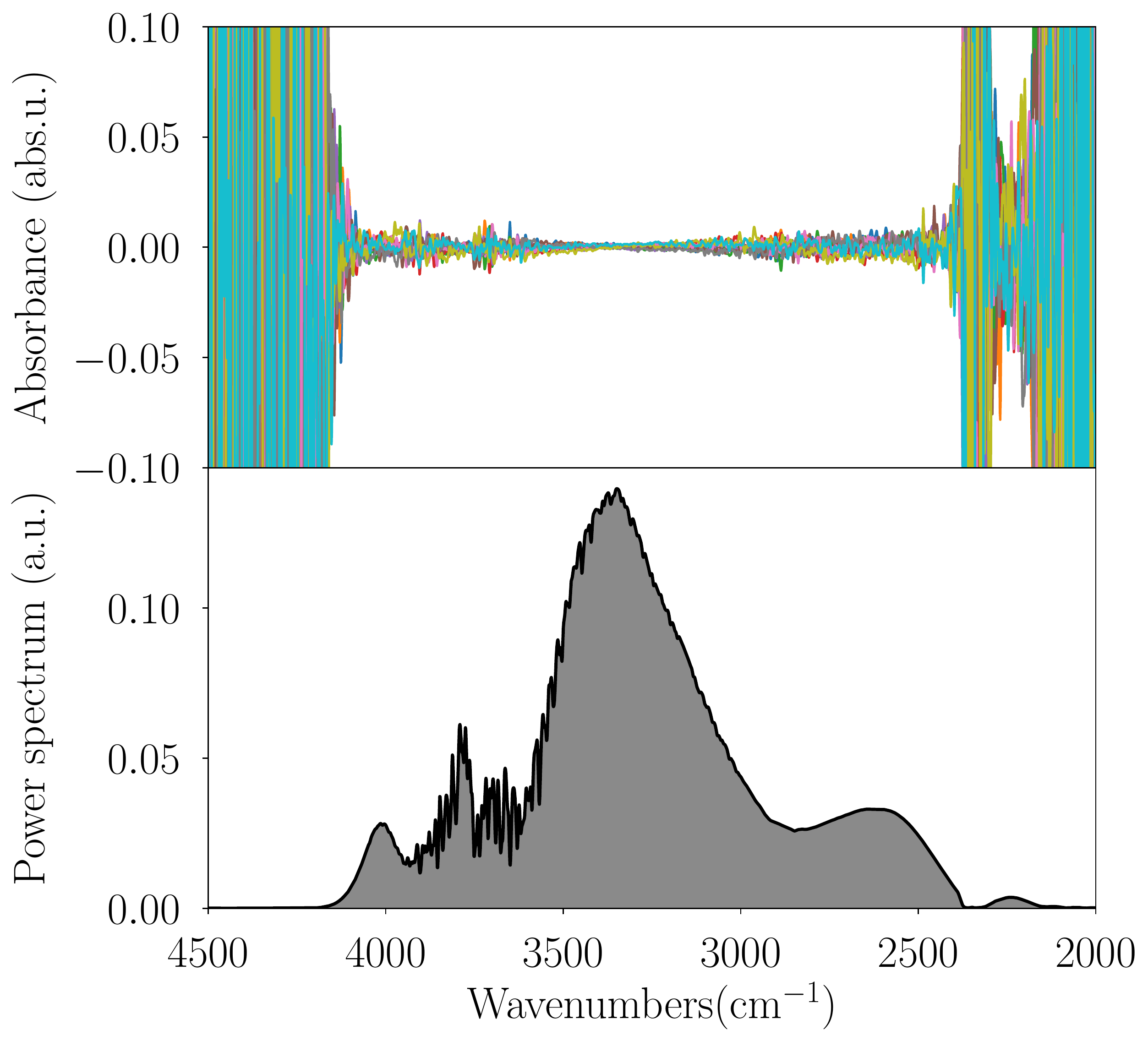}
  \caption{Thorlabs SC4500 InF\textsubscript{3} fiber based supercontinuum source (300~mW, 50~MHz); RMS=0.0024 (abs.u.)}\label{fig:thorlabs100lines}
 \end{subfigure}
  \begin{subfigure}[t]{.45\columnwidth}
  \includegraphics[width=\columnwidth]{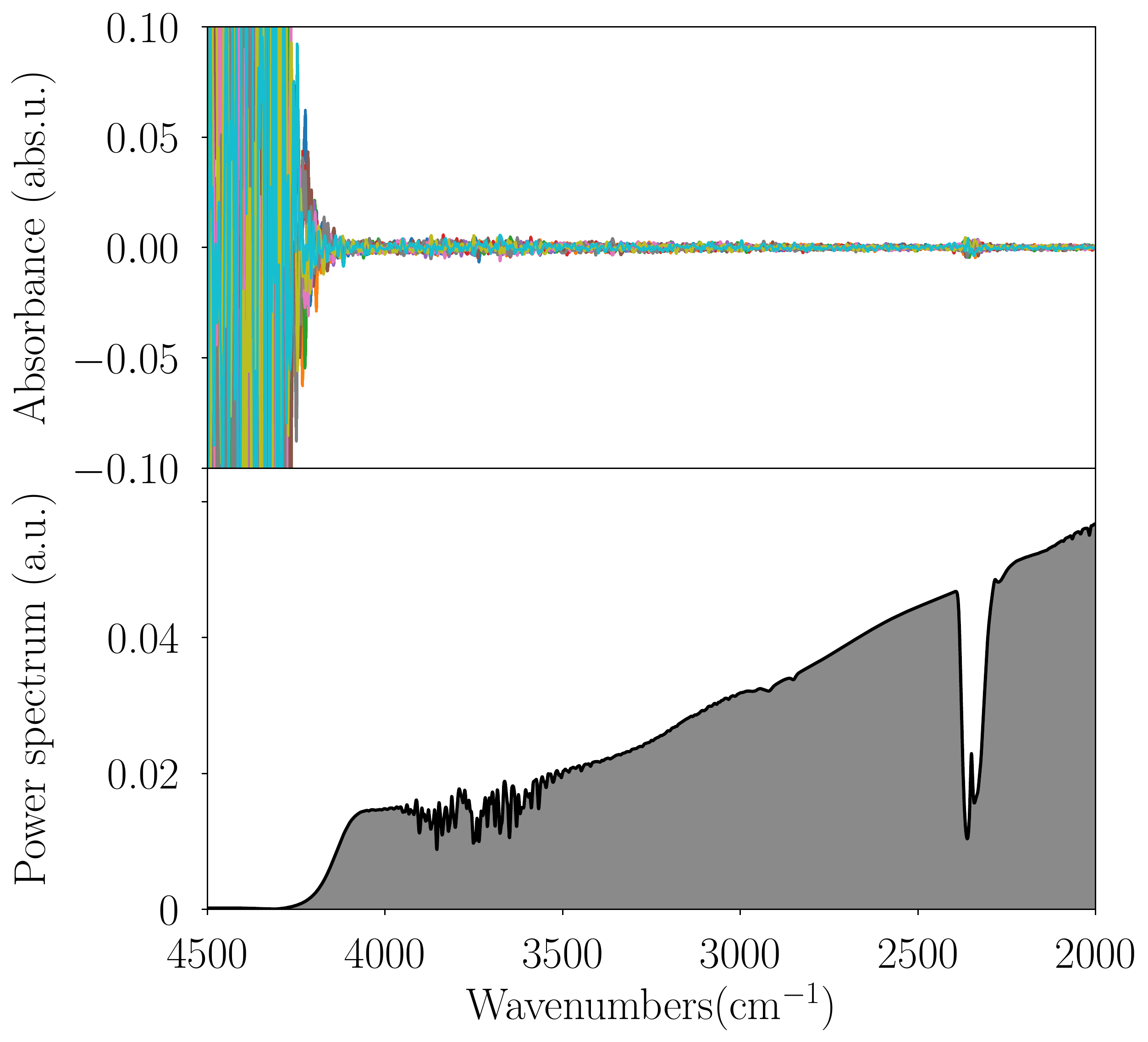}
  \caption{Standard on-board thermal emitter; \\RMS=0.0012 (abs.u.)}
 \end{subfigure}
  \caption{Zero absorbance lines measured using a standard FTIR instrument employing (a-c) mid-IR supercontinuum sources and (d) a conventional thermal emitter; emission power spectra are shown below in corresponding sub-plots for reference (characteristic absorptions of atmospheric CO\textsubscript{2} and H\textsubscript{2}O are present), detector dark noise floor is 3.3$\times$10\textsuperscript{-5} (a.u); the FTIR spectrometer, detector (MCT) and their parameters were set the same for all measurements, thus, the measurement time for noise evaluation was set constant (around 5~sec per spectrum); the mid-IR part of the spectrum was selected using a 2.4~\textmu m edge-pass spectral filter; the RMS errors are calculated for the spectral bandwidth from 4025~cm\textsuperscript{-1} to 2280~cm\textsuperscript{-1} [to 2500~~cm\textsuperscript{-1} for (c) due to the limited coverage].}
 \begin{tikzpicture}[overlay,thick]
 \end{tikzpicture}
 \label{fig:100lines}
\end{figure}

In order to characterize the spectral performance of mid-IR supercontinuum sources and to compare them to a standard thermal emitter, excluding possible contributions from the instrumental noise, we used a commercial FTIR spectrometer (Bruker Optics, Vertex 70) as the core system and a mid-IR detector (Mercury Cadmium Telluride variable gap detector, MCT, Vigo PCI-4TE-12, detectivity
D$~=~2.0\times~10^{9}$~$\mathrm{cm\cdot \sqrt{Hz}\cdot W^{-1}}$) for all characterization measurements. The acquisition parameters and settings of the spectrometer and detector were fixed, so the measurement time for noise estimation was set constant (around 5~sec per spectrum) for different sources. Figure~\ref{fig:100lines} depicts the zero absorbance lines measured for three mid-IR supercontinuum sources and for the built-in FTIR Globar respectively.

Since supercontinuum sources are pulsed low-duty cycle light emitters, a boxcar integrator (gated integrator, Zurich Instruments, UHFLI) was used to demodulate and pre-process the signal for acquisition by the read-out electronics of the spectrometer. Therefore, the inter-pulse noise was excluded by averaging the signal during an optimally configured time gate (from 20 to 50~ns). This type of integrator was selected as it is more efficient for non-sinusoidal low-duty cycle sources compared to lock-in amplifiers\cite{doi:10.1063/1.5022266}. We assumed that the integrator does not introduce significant noise, allowing to compare the pulsed supercontinua with the CW thermal emission.

All the measurements shown in Fig.~\ref{fig:100lines} were performed with a locked configuration and parameters of the spectrometer. An edge-pass filter (2.4~\textmu m cut-on wavelength) was used to select the mid-IR spectral band. In order to operate the detector close to the saturation regime, the size of the aperture was adjusted due to the different brightness levels of the light sources. Since one of the supercontinuum lasers (Leukos, InF\textsubscript{3} fiber based) has considerably higher pulse energy (650~mW average power, 250~kHz repetition rate) compared to the other sources, a strong neutral density filter was employed to scale the emission of the former (1\% transmission, Thorlabs NDIR20B, spectral intensity noise is assumed to be scaled proportionally). The mirror frequency of the FTIR system was fixed at 1~kHz for the characterization, thus, the measurement time window was constant for noise evaluation. Depending on the pulse repetition rate of the source, the boxcar integrator (i.e. an interferogram demodulation device) was configured individually in order to avoid a smear out of the high frequency components in the respective interferograms. This smear out is caused due to the high mirror velocity relative to the effective bandwidth of the boxcar integrator. The boxcar time constant was set to 1024~cycles (equivalent to 410~\textmu s time constant) for the supercontinuum source from NKT Photonics, to 128~cycles (512~\textmu s time constant) for the supercontinuum laser from Leukos, and to 32768~cycles (655~\textmu s time constant) for Thorlabs SC4500. 
The obtained spectra were smooth owing to the high signal-to-noise ratio.

Considering the results of the characterization depicted in Fig.~\ref{fig:100lines} and the obtained RMS errors, it has to be noted that the high repetition rate supercontinuum generators from NKT Photonics and Thorlabs possess the superior spectral stability (spectral deviations cannot be distinguished by eye), and in the small ranges they surpassed the thermal emitter in the given configuration. However, the spectral noise performance of the ZBLAN supercontinuum-based FTIR can still be improved by flattening the spectral shape or removing the intense spectral peak to efficiently exploit the dynamic range of the detector and DAQ unit, which can result in comparable noise levels in relatively broad spectral ranges.

All the characterized supercontinuum sources are conventional, based on an anomalous dispersion generation scheme. Figure~\ref{fig:100lines} confirms and illustrates the commonly known fact that supercontinuum sources employing pumping in the anomalous dispersion regime exhibit relatively high fluctuations (especially in comparison to the novel concept of all-normal dispersion supercontinuum generators\cite{PhysRevLett.90.113904,Moller:12,klimczak_direct_2016,Heidt:17}). However, the figure also shows the importance of pulse averaging, so that high pulse repetition rate emitters (in MHz or tens MHz range) can diminish the effects of pulse-to-pulse spectral instabilities for most routine mid-IR spectroscopic applications. An illustrative example for noise reduction by integration using high repetition rates in the extreme noise sensitive technique of spectral-domain optical coherence tomography can be found in\cite{Maria:17}. Moreover, we would like to point out that the enhancement in spectral brightness enables to significantly extend light-matter interaction path lengths and to probe more molecules (i.e. intensifying absorbance signals with respect to noise). Thus, the relatively higher spectral instabilities (in the standard measurement time scale) of supercontinuum sources compared to standard thermal emitters become insignificant in practice.

\begin{figure}[ht]
\centering
  \begin{tikzpicture}
\centering
 \begin{scope}[x={(image.south east)},y={(image.north west)}]
    \node at (0.5,0.5) { \includegraphics[width=0.75\columnwidth]{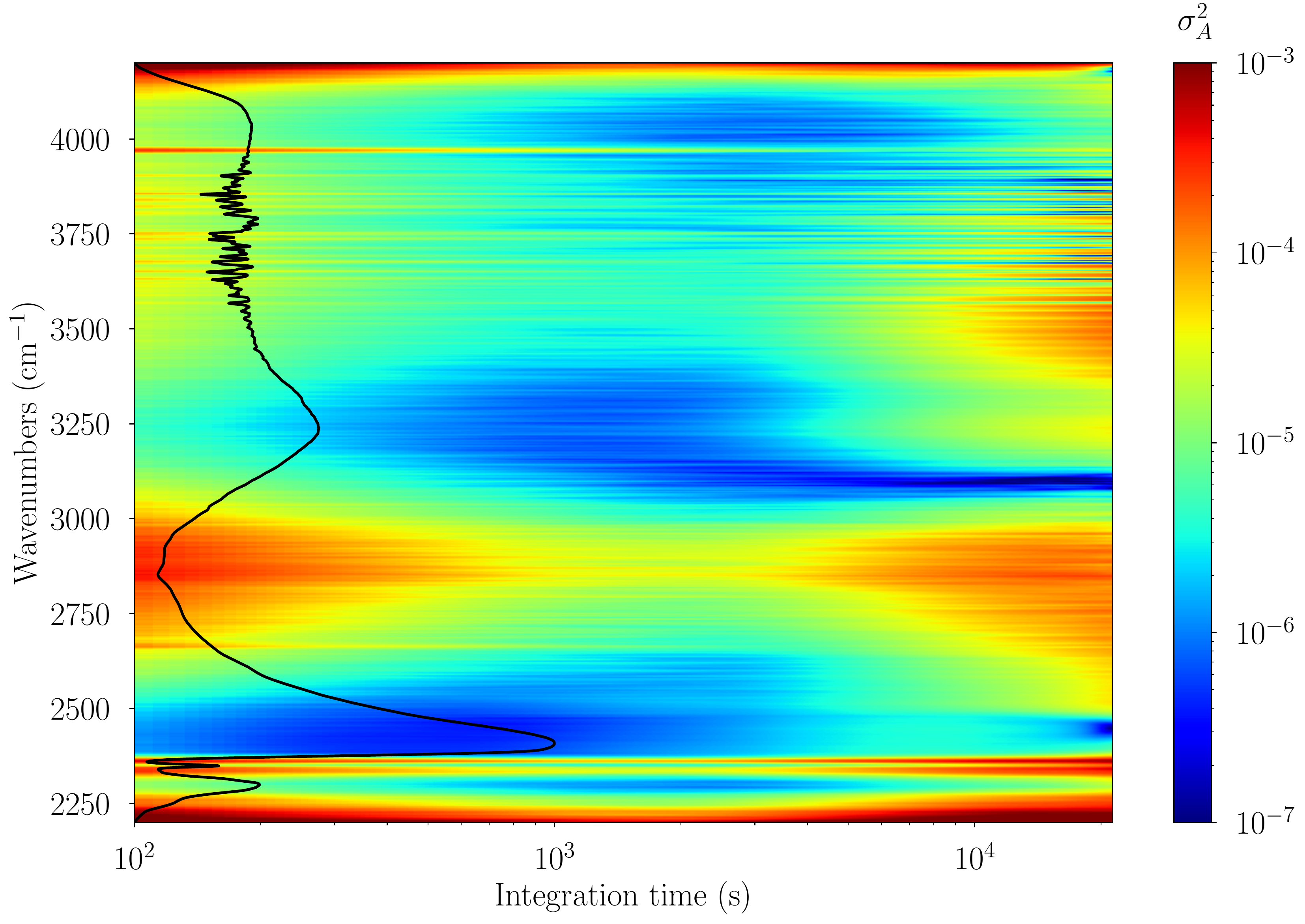}};
 \end{scope}
\end{tikzpicture}
  \caption{Spectral 2-dimensional Allan variance for the emission of the mid-IR supercontinuum source (NKT Photonics, SuperK mid-IR) visualizing relative spectral drifts (an FTIR instrument was used for evaluation); this is a source-specific characterization measurement showing the spectral dependent stability and drifts; for small cluster sizes it correlates with zero absorbance lines but provides additional insights into temporal behavior; the emission spectrum is shown for reference as the black overlay.}
 \label{fig:2d-allan-plot}
\end{figure}

Additionally, we extend the concept of the Allan variance defined in the previous section and apply it to visualize the presence of spectrally dependent drifts for the FTIR-based instrument (see Fig.~\ref{fig:2d-allan-plot}). It should be noted, however, that the resulting 2-dimensional Allan variance should not be considered as an absolute assessment for the emitter stability (i.e. for evaluation of absolute optimal integration time), but rather as a relative one (illustrating relative behavior of constituent spectral components). Contributions from other sub-systems (such as detector instabilities, drifts of the power supplies, temperature-induced variations, deviations in the spectrometer parameters etc.) are superimposed on the drifts of the mid-IR supercontinuum laser source itself. For instance, strong and sharp changes in the integral of the recorded emission spectra were observed in the mornings due to changes in the building's electrical power supply, although, these changes are not visible in the average power stability characterization of the emission (i.e. more isolated source characterizations). Therefore, we neglected the above-mentioned non-emission related drifts and calculated the Allan variance for a 12 hour time period where those drifts did not occur (i.e. source related drifts presumably dominate). Hence, any spectral instability for this time period can be accounted to be a property of the analyzed laser and not of the aforementioned external sources. Thus, the practical reasoning behind this analysis to distinguish the relative time-stability of the different spectral regions of the supercontinuum emission is valid.

The intensity of various spectral components were analyzed over the time interval of 12~hours, which is also a reasonable time-span for single-day measurements. The sampling period for the emission spectra was around 100 sec, the spectral resolution was set to 4~cm\textsuperscript{-1}. Figure~\ref{fig:2d-allan-plot} depicts the resulting 2-dimensional Allan variance. It can be noted that the initial shape (i.e. the variance for single-measurement clusters) of the spectral Allan variance correlates with the zero absorbance lines shown in Fig.~\ref{fig:nkt100lines}. However, in contrast to zero-absorbance lines, which always characterize the deviation between two subsequent spectra (only short-time changes can be detected), the Allan variance provides the assessments for statistically average spectral stability across the entire time interval, regardless of the initial time point.

Analyzing the Allan plot shown in Fig.~\ref{fig:2d-allan-plot}, several pronounced features can be distinguished. For instance, the broad and strong spectral peak centered at around 2400~cm\textsuperscript{-1}, which appeared to be highly stable in the short run (superior to the thermal emitter, as also demonstrated by zero absorbance lines), exhibits the presence of relatively stronger intensity drifts and, therefore, lower integration capabilities than other spectral sub-bands (the optimal integration time here is below 1000~sec). For the rest of the spectrum, the slope characteristic for white noise dominates until reaching the minimum of Allan variance. The mean optimal averaging time is in the range of around 1500~sec, which is in agreement with the measurements shown in the previous section [Fig.~\ref{fig:allan}]. Furthermore, for some spectral regions a noise reduction after 10\textsuperscript{4}~sec can be observed. These facts confirm our assumptions that the selected time interval of 12~hours is suitable for the analysis of the emission stability. A particularly interesting spectral behavior, i.e., an increased long-term stability of the spectral shape with a predominance of white noise, can be also observed in the range between approximately 3000~cm\textsuperscript{-1} and 3250~cm\textsuperscript{-1}. This spectral window (especially around 3100~\textmu m) is quite important for mid-IR spectroscopy due to the presence of distinct absorption bands (stretching vibrations of the C-H, N-H, O-H groups). 

\subsection{Summary of the properties and their relation to specific spectroscopic applications}

The purpose of this section is to generalize the most important and, at the same time, the most distinctive properties of supercontinuum laser sources relevant for applied mid-IR spectroscopy.
Therefore, we conclude this part by summarizing and listing the characteristic parameters and identifying potential applications that promise to be or have already been advanced by mid-IR supercontinuum technology.

\begin{figure}[ht]
\centering
\begin{tikzpicture}
\centering
 \begin{scope}[x={(image.south east)},y={(image.north west)}]
    \node at (0.5,0.5) {\includegraphics[width=0.85\columnwidth]{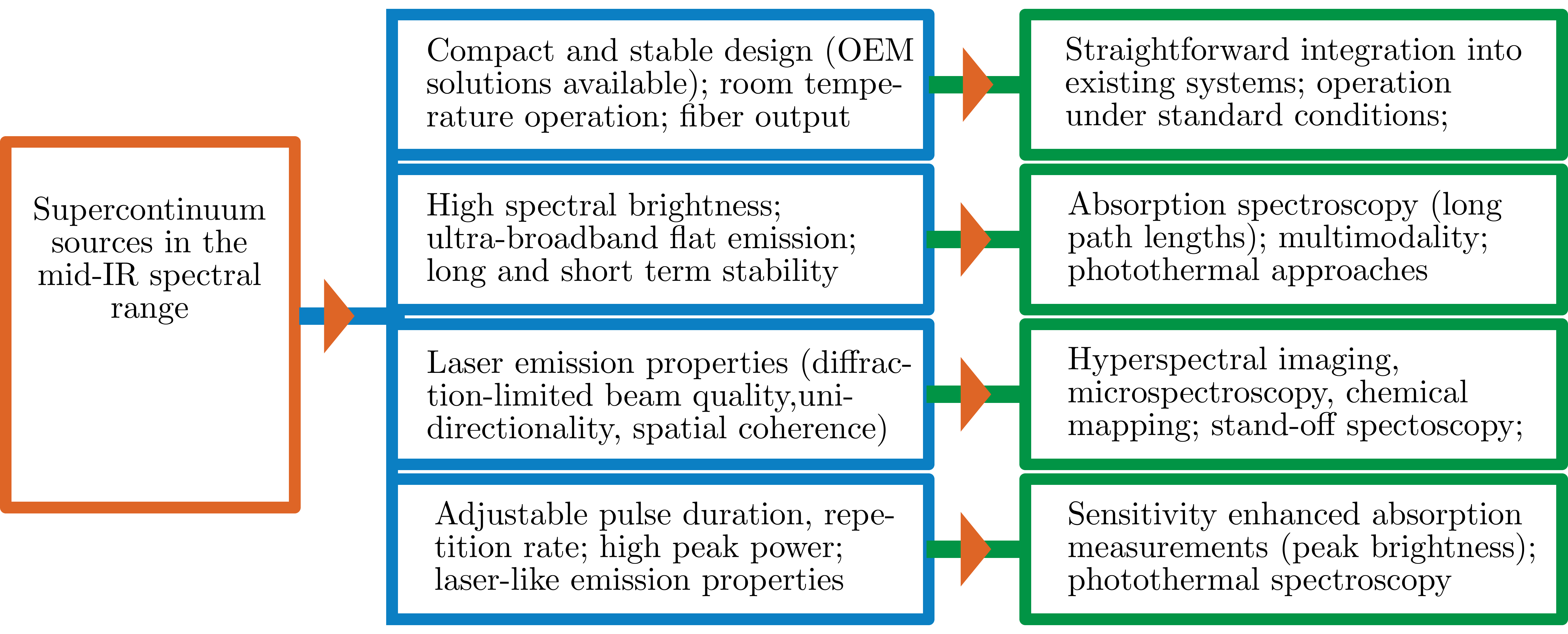}};
     \node at (0.157,0.38) {\includegraphics[scale=0.058]{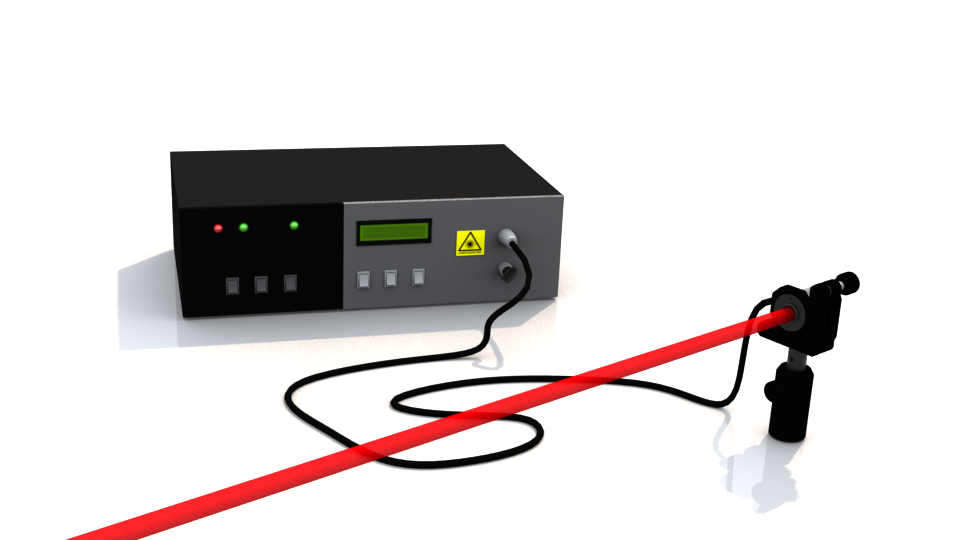}};
 \end{scope}
\end{tikzpicture}
 \caption{Brief summary of the characteristic features of mid-IR supercontinuum laser sources; highlighted according to potential applied cases and scenarios.}
 \label{fig:props}
\end{figure}

The brief summary of the key properties and their relation to certain applied scenarios, in which they are expected to have the strongest impact, is given in Fig.~\ref{fig:props}. Here we once again emphasize the importance of laser-like emission properties of supercontinua (beam quality, spatial coherence, uni-directionality) coupled with the high spectral brightness, stability, and ultra-broadband spectral coverage. The mid-IR supercontinuum technology can provide instant access to the functional group as well as the fingerprint area with a single emitter. Thus, such a combination of emission properties is unique and non-specific to any other mid-IR laser source. Hence, the strict categorization given here is simplified to show the most striking implications. Discussion of the specific properties of supercontinuum radiation (no mid-IR supercontinuum considered at this point) and their relation to spectroscopic application scenarios can also be found in\cite{kaminski_supercontinuum_2008}. 

In the following section, we supplement the scheme reviewing the current state-of-the-art and providing up-to-date achievements in the distinguished fields.

\section{Supercontinuum sources in mid-IR spectroscopy: an overview of state-of-the-art applications}

Supercontinuum laser sources have been used for applied IR spectroscopy (primarily near-IR) for more than a decade. However, they are now at the stage of widespread adoption, becoming standard laboratory equipment that can advance well-established spectroscopic techniques and methods. In this section, we review the current state of supercontinuum-based mid-IR spectroscopy and list the most valuable reports in categorised tables (see Tables \ref{tab:table1}, \ref{tab:table2}, and \ref{tab:table3}). The topical tables contain a summary of the results and provide relevant information about the application scenario and the source used; emission spectral ranges are indicated in brackets.\ignore{ (shown in brackets) and the range where measurements were carried out (listed below).}

\arrayrulecolor[HTML]{000000}
\begin{footnotesize}
\begin{longtable}{>{\hspace{0pt}}m{0.02\columnwidth}>{\raggedright\hspace{0pt}}m{0.22\columnwidth}>{\raggedright\hspace{0pt}}m{0.22\columnwidth}>{\raggedright\hspace{0pt}}m{0.22\columnwidth}>{\hspace{0pt}}m{0.13\columnwidth}}
\caption{Overview of mid-IR supercontinuum applications in direct absorption spectroscopy.\label{tab:table1}}\\ 
\toprule
\multicolumn{1}{>{\hspace{0pt}}m{0.02\columnwidth}}{\texttt{\#}} & \multicolumn{1}{>{\centering\hspace{0pt}}m{0.22\columnwidth}}{Type of supercontinuum} & \multicolumn{1}{>{\centering\hspace{0pt}}m{0.22\columnwidth}}{Application description} & \multicolumn{1}{>{\centering\hspace{0pt}}m{0.22\columnwidth}}{Details} & \multicolumn{1}{>{\centering\hspace{0pt}}m{0.13\columnwidth}}{Ref. Year} \endfirsthead 
\midrule
1 & Step-index ZBLAN fiber based, custom, same as in\cite{4811103}; varying repetition rate in MHz and sub-MHz range, 10.5~W average power (reduced to 1.5~W for the spectroscopic measurements); (12500-2220~cm\textsuperscript{-1}) & Identification and differential damage \textit{in vitro} of lipids and proteins; absorption spectroscopy of the constituents of normal artery and atherosclerotic plaque;& Measurements performed within C-H fatty acid and cholesterol esters absorption bands; results compared with thermal- and synchrotron-based systems; adequate signal-to-noise ratio achieved & \makecell{ Ke et al.\\ \cite{Ke:09} 2009}\\ 
\hline
\midrule
2 & Step-index ZBLAN fiber based, custom; 10~kHz repetition rate, 490 mW average power; (5000-2860~cm\textsuperscript{-1})  & Qualitative and quantitative supercontinuum laser absorption spectroscopy for a multi-component atmospheric gas mixture & Identification and concentration estimation of
an unknown multi-component atmospheric gas-mixture (demonstrated for methane) & \makecell{ Cezard et al.\\ \cite{10.1117/12.898227} 2011}\\ 
\midrule
3 & Step-index ZBLAN fiber based, commercial; variable repetition rate between 30 and 50 kHz, 75 mW average power; (5900-2400~cm\textsuperscript{-1}) & Absorption measurements of oils, polymers, aqueous solutions of acetic acid, ethyl-2-cyanoacrylate in transmission and reflection geometry & Noise and polarization characterized; application in various scenarios demonstrated; advantages and potential applications defined & \makecell{ Kilgus et al.\\ \cite{10.1117/12.2225886} 2016}\\ 
\midrule
4 & Step-index ZBLAN fiber based, commercial; 35~kHz repetition rate, 60~mW average power ; (8335-2380~cm\textsuperscript{-1})   & Transmission measurements of barley endosperm and barley oil with further classification  & Potentials of supercontinuum-based system for analysis of food samples demonstrated;  technical issue of the spectra distortion due to the absence of pulse demodulation put forward & \makecell{ Ringsted et al.\\  \cite{doi:10.1177/0003702816652361} 2016}\\ 
\midrule
5 & Step-index ZBLAN fiber based, custom\ignore{(7~m, ca. 9~\textmu m core size)}; 100~kHz repetition rate, 160~mW average power; (11100-2700~cm\textsuperscript{-1})  & \raggedright Gas detection and concentration measurements for acetylene and methane [see Fig.~\ref{fig:amiot_system}]  & \raggedright Cavity enhanced spectroscopy for multi-component gas detection with sub-ppm accuracy demonstrated [see Fig.~\ref{fig:amiot_calibration_curves}] & \makecell{ Amiot et al.\\ \cite{doi:10.1063/1.4985263} 2017}\\ 
\midrule
6 & Step-index ZBLAN fiber based, commercial; 40~kHz repetition rate, 75 mW average power; (5715-2380~cm\textsuperscript{-1})  & Multi-bounce ATR spectroscopy, quantification of hydrogen peroxide aqueous solutions  & 3-times improved detection limits compared to thermal emitters demonstrated & \makecell{ Gasser et al.\\ \cite{Gasser:18} 2018}\\ 
\midrule
7 & Step-index ZBLAN fiber based, commercial;  repetition rate 2.5~MHz, 500~mW average power; (6665-2380~cm\textsuperscript{-1}) & Multi-species gas sensing based on supercontinuum source and upconversion detection & System for multi-species gas detection of broadly absorbing gas species (nitrous oxide, ethylene, methane, ethane, acetaldehyde) demonstrated; high speed (~20 ms) sensing of ethane with ~15 ppbv$\cdot$s\textsuperscript{-1} sensitivity achieved &  \makecell{ Jahromi et al.\\ \cite{Jahromi:19} 2019}\\ 
\hline
\newpage
\texttt{\#} &\centering Type of supercontinuum& \centering Application description&\centering Details&{\hspace{9pt} Ref. Year} \\ 
\midrule
8 & Step-index ZBLAN fiber based, commercial;  repetition rate 2.5~MHz, above 450~mW average power; (6665-2380~cm\textsuperscript{-1}) & Detection of multiple broadband absorbing gas species with sensitivity in the sub-ppmv level & Importance of brightness highlighted;  application for quality control of stored fruits demonstrated including simultaneous detection of ethylene, ethanol, ethyl acetate, acetaldehyde, methanol, acetone, and water&  \makecell{ Jahromi et al.\\ \cite{s19102334} 2019} \\ 
\midrule
9 & Step-index ZBLAN fiber based, commercial;  repetition rate adjusted to 3~MHz, 475~mW average power; (9090-2270~cm\textsuperscript{-1})  & Sensitivity-enhanced supercontinuum-based FTIR instrument demonstrated, compared to the conventional system
& 4-times enhanced detectivity demonstrated through the extension of interaction path length, exemplified for aqueous formaldehyde series; detection method based on lock-in amplification proposed and discussed & \makecell{ Zorin et al.\\ \cite{Zorin:AS:20} 2020}  \\ 
\midrule
10 & Step-index ZBLAN fiber based, commercial;  repetition rate 2.5~MHz, 500~mW average power; (6665-2380~cm\textsuperscript{-1}) & Multi-species trace gas sensing based on a high-repetition-rate supercontinuum source and a scanning grating spectrometer (30~m multipass absorption cell used) & Importance of pulse integration (using lock-in demodulation) demonstrated, detection limit in the order of 100 ppbv~Hz\textsuperscript{-1/2} for various hydrocarbons, alcohols, and aldehydes demonstrated &  \makecell{ Jahromi et al.\\ \cite{EslamiJahromi:20} 2020} \\ 
\midrule
11 & Step-index tellurite fiber based, custom; 80~MHz repetition rate, 40~mW average power; (5000-2855~cm\textsuperscript{-1}) & Gas sensing based on a high-repetition-rate mid-IR supercontinuum laser source; feasibility of multiple detection of greenhouse gases demonstrated & Detection and analysis of carbon dioxide and methane and their binary mixture using compact multipass cell& \makecell{ Lemi{\`{e}}re et al.\\ \cite{Lemi_re_2021} 2021}\\ 
\midrule
12 & Step-index ZBLAN fiber based, commercial;  repetition rate 2.5~MHz, 500~mW average power; (5000-2500~cm\textsuperscript{-1}) & Multi-species trace gas sensing (sub-ppmv Hz\textsuperscript{-1/2} sensitivity) using a mid-IR supercontinuum source and a Fourier transform spectrometer & Fast, sensitive and high resolution trace gas sensor for fruit quality monitoring demonstrated; balanced detection scheme employed; concentration measurements (ethanol, acetaldehyde, ethyl acetate, ethylene, acetone, methanol) &  \makecell{ Jahromi et al.\\ \cite{Jahromi:21} 2021} \\ 
\midrule
13 & 1. Step-index ZBLAN fiber based, commercial;  repetition rate 2.5~MHz, 450~mW average power; (5000-2500~cm\textsuperscript{-1}) \newline 2. Chalcogenide fiber based, custom (DTU Fotonik)\cite{Woyessa:21}; repetition rate 3~MHz, 86~mW average power; (6660-950~cm\textsuperscript{-1}) & Fast-scanning Fourier transform spectrometer based on high-repetition-rate mid-IR supercontinuum sources for trace gas detection (demonstrated for ethyl acetate, ethane, nitrous oxide, sulfur dioxide)  & Trace gas detection (spectral resolution of 750 MHz) using supercontinuum sources beyond 5~\textmu m wavelength; noise of various laser-based 
spectroscopy methods compared; supercontinuum as a alternative for real-life applications introduced &  \makecell{ Abbas et al.\\ \cite{Abbas:21} 2021} \\ 
\multicolumn{1}{>{\hspace{0pt}}m{0.02\columnwidth}}{} &  \multicolumn{1}{>{\hspace{0pt}}m{0.22\columnwidth}}{} & \multicolumn{1}{>{\hspace{0pt}}m{0.22\columnwidth}}{} & \multicolumn{1}{>{\hspace{0pt}}m{0.22\columnwidth}}{} & \multicolumn{1}{>{\hspace{0pt}}m{0.13\columnwidth}}{} \\
\bottomrule
\end{longtable}
\end{footnotesize}

\begin{figure}[hb]
\centering
 \begin{subfigure}[t]{.42\columnwidth}
 \includegraphics[width=\columnwidth]{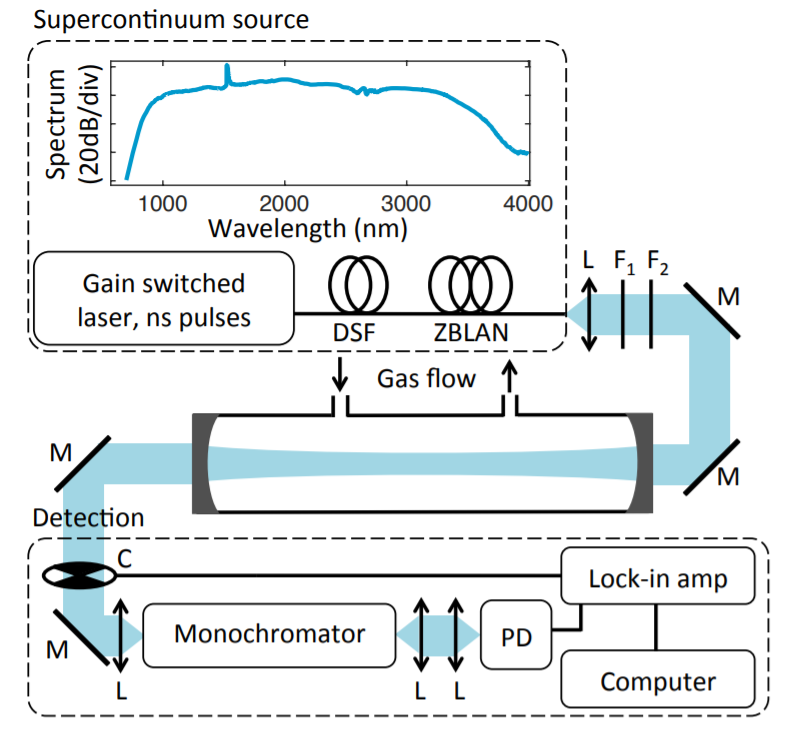} 
 \caption{Cavity-enhanced supercontinuum-based absorption spectroscopy system}\label{fig:amiot_system}
 \end{subfigure}
 \begin{subfigure}[t]{.42\columnwidth}
  \includegraphics[width=\columnwidth]{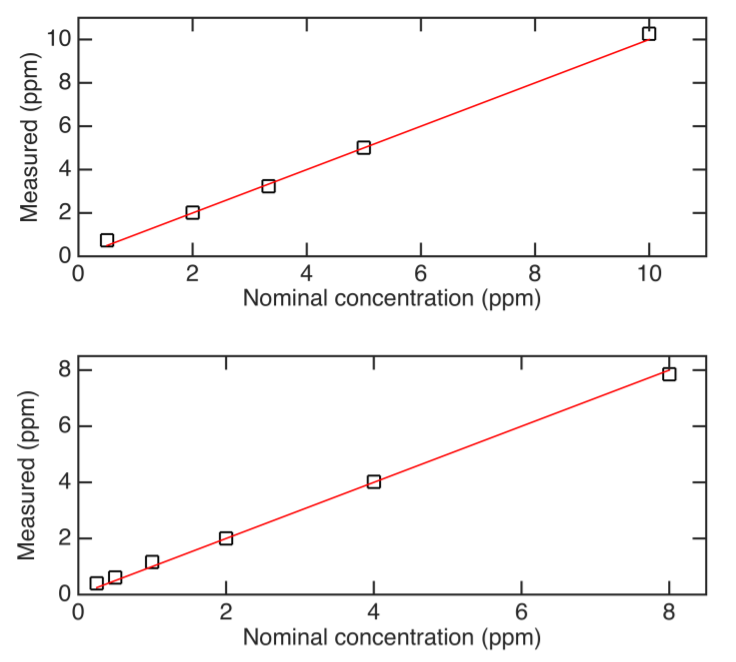}
  \caption{Calibration curves for acetylene (top) and methane (bottom) concentration series}\label{fig:amiot_calibration_curves}
 \end{subfigure}
  \caption{Broadband supercontinuum-based cavity enhanced absorption spectroscopy in
the mid-IR spectral range\cite{doi:10.1063/1.4985263}, reproduced with permission from AIP Publishing.}
 \begin{tikzpicture}[overlay,thick]
 \end{tikzpicture}
 \label{fig:amiot}
\end{figure}

\newpage
Figure~\ref{fig:amiot} reproduces recently published and illustrative results on high-resolution, high-sensitivity  (sub-ppm for acetylene and methane) cavity enhanced absorption spectroscopy for gas measurements. An experimental system incorporating a monochromator and a mid-IR supercontinuum laser source is shown in\ref{fig:amiot_system}.

It has to be noted that in photoacoustic mid-IR spectroscopy till now only one remarkable research based on the FTIR-photoacoustic approach has been reported\cite{Mikkonen:18} (Mikkonen et al., 2018). The employed source was ZBLAN fiber based, same as in \cite{doi:10.1063/1.4985263} (repetition rate adjusted depending on the sample, from 70~kHz to 400~kHz); the emission range is from 12500~cm\textsuperscript{-1} to 2703~cm\textsuperscript{-1} (measurements were performed in two sub-bands, 5715-5000~cm\textsuperscript{-1} and 3175-2860~cm\textsuperscript{-1}). The research demonstrated cantilever-enhanced photoacoustic spectroscopy for broadband gas sensing. Absorption measurements of water vapor and methane with significant signal enhancements (by a factor of 70) compared to the conventional system based on a thermal emitter have been demonstrated.

\begin{footnotesize}
\begin{longtable}{>{\hspace{0pt}}m{0.02\columnwidth}>{\raggedright\hspace{0pt}}m{0.22\columnwidth}>{\raggedright\hspace{0pt}}m{0.22\columnwidth}>{\raggedright\hspace{0pt}}m{0.22\columnwidth}>{\hspace{0pt}}m{0.13\columnwidth}}
\caption{Overview of supercontinuum-based mid-IR hyperspectral imaging and microspectroscopy.\label{tab:table2}}\\ 
\toprule
\multicolumn{1}{>{\hspace{0pt}}m{0.02\columnwidth}}{\texttt{\#}} & \multicolumn{1}{>{\centering\hspace{0pt}}m{0.22\columnwidth}}{Type of supercontinuum} & \multicolumn{1}{>{\centering\hspace{0pt}}m{0.22\columnwidth}}{Application description} & \multicolumn{1}{>{\centering\hspace{0pt}}m{0.22\columnwidth}}{Details} & \multicolumn{1}{>{\centering\hspace{0pt}}m{0.13\columnwidth}}{Ref. Year} \endfirsthead 
\midrule
1 & Step-index ZBLAN fiber based, custom; 40 MHz repetition rate, spectral brightness of 300 kW/(nm$\cdot$m\textsuperscript{2}$\cdot$sr) at 2857~cm\textsuperscript{-1} given; (7145-2500~cm\textsuperscript{-1}) & Hyperspectral imager based on a bright supercontinuum source and a monochromator (PbSe detector used) (transmission, mapping) & Importance of spectral brightness discussed; microspectroscopy demonstrated, exemplified for hyperspectral imaging of oil/water mixture; 2~nm spectral resolution, spatial resolution of around 20~\textmu m verified&\makecell{ Dupont et al.\\ \cite{Dupont:12} 2012 }\\ 
\midrule
2 & Step-index ZBLAN fiber based; repetition rate not stated, 1.2~W average power (0.9 mW at sample); (5000-2220~cm\textsuperscript{-1}) &  Hyperspectral imaging using bright supercontinuum source, exemplified for polymer film imaging (transmission, full-field) & Prototype of hyperspectral imager based on acousto-optic tunable filter and InSb camera demonstrated (5~\textmu m resolution stated)&\makecell{ Farries et al.\\ \cite{Farries_2015} 2015 }\\ 
\midrule
\newpage
\texttt{\#} &\centering Type of supercontinuum& \centering Application description&\centering Details&{\hspace{9pt} Ref. Year} \\ 
\midrule
3 & Step-index ZBLAN fiber based, custom; 30-350 kHz repetition rate, >800~mW average power (set to 30~kHz for imaging); (5000-2220~cm\textsuperscript{-1}) & Hyperspectral microscopy based on acousto-optic tunable filter and thermal camera (transmission, full-field) & Performances of QCLs, thermal emitters, and synchrotron radiation for microspectroscopy discussed; imaging system and its performances for polystyrene samples demonstrated (3.5 cm\textsuperscript{-1} resolution)&\makecell{ Lindsay et al.\\ \cite{10.1117/12.2210836} 2016}\\ 
\midrule
4 & Step-index ZBLAN fiber based, custom; 30-300 kHz repetition rate, 1.1~W average power (set to 30~kHz and 50~mW for imaging); (5000-2200~cm\textsuperscript{-1})  & Hyperspectral imaging system for rapid assessment of cells for cytological diagnosis based on acousto-optic tunable filter and thermal camera (transmission, full-field) & Exemplified for colon cells; high spatial resolution and high speed imaging compared with the FTIR system demonstrated; potentials for point of care screening on live patient highlighted & \makecell{ Farries et al.\\ \cite{10.1117/12.2250811} 2017 }\\ 
\midrule
5 & Step-index ZBLAN fiber based, commercial;  repetition rate 40~kHz, 100~mW average power; (5550-2220~cm\textsuperscript{-1}) &  Broadband upconversion-based imaging,  proof-of-principle (transmission, full-field) & Prototype of low-NA (aberrated due to parabolic mirror) hyperspectral upconversion imager demonstrated; no spectral diversification given, but indirectly shown in images (phase-matching distribution) & \makecell{ Huot et al.\\ \cite{10.1117/12.2251805} 2017 }\\ 
\midrule
6 & Chalcogenide fiber based (tapered large-mode-area photonic crystal fiber from highly purified Ge\textsubscript{10}As\textsubscript{22}Se\textsubscript{68} glass), custom; 21 MHz repetition rate, 25~mW average power (see~\cite{Petersen:17}); (5000-1335~cm\textsuperscript{-1})  & Multispectral tissue imaging of nontumoral colon tissue section in the diagnostically important fingerprint region (transmission, mapping) & Proof-of-principle demonstration of mid-IR multispectral microscopic imaging; compared to conventional FTIR system; relatively low-noise images obtained, 12.4~\textmu m spatial resolution (at 1667 cm\textsuperscript{-1}) stated&\makecell{ Petersen et al.\\ \cite{Petersen:18} 2018 }\\ 
\midrule
7 & Step-index ZBLAN fiber based, custom\ignore{(7~m, ca. 9~\textmu m core size)}; 4.15~MHz repetition rate of the pulse train, several satts of average power (reduced by neutral density and band-pass filters); (5000-2665~cm\textsuperscript{-1})  & Fourier transform infrared microspectroscopy (transflection, mapping)  & Chemical mapping of lipid vesicles in a liver demonstrated; compared with synchrotron- and thermal-based instruments \textendash improved signal-to-noise ratio/shorter acquisition times achieved; 3~\textmu m spatial resolution (at around 2950 cm\textsuperscript{-1}) stated &\makecell{ Borondics et al.\\ \cite{Borondics:18} 2018 }\\ 
\midrule
8 & Step-index ZBLAN fiber based, commercial;  repetition rate 2.5~MHz, 500~mW average power; (6450-2220~cm\textsuperscript{-1})  & Diffraction-limited chemical mapping (reflection, spatial resolution of 2.8~\textmu m at 2700~cm\textsuperscript{-1}) based on MEMS Fabry-P\'erot filter spectrometer & Diffraction-limited microspectroscopy verified; imaging of red blood cells (spectrum of a single cell recorded, see details in Fig.~\ref{fig:jakob_blood}) and various polymers demonstrated; sensitivity advantages over standard equipment discussed &\makecell{ Kilgus et al.\\ \cite{Kilgus:18} 2018 }\\ 
\midrule
\newpage
\texttt{\#} &\centering Type of supercontinuum& \centering Application description&\centering Details&{\hspace{9pt} Ref. Year} \\ 
\midrule
9 &  Step-index ZBLAN fiber based, commercial;  repetition rate 2.5~MHz, 490~mW average power; (9090-2275~cm\textsuperscript{-1}) & Multimodal hyperspectral and structural imaging for non-destructive testing in art diagnosis (reflection, mapping) & Active hyperspectral imager combined with 3D-tomographic modality demonstrated; performance for imaging of art mock-ups (oil paintings) reported & \makecell{ Zorin et al.\\ \cite{10.1117/12.2528279} 2019 }\\ 
\midrule
10 & Step-index ZBLAN fiber based, commercial;  repetition rate 2.5~MHz, 490~mW average power; same as in~\cite{10.1117/12.2528279}; (9090-2275~cm\textsuperscript{-1})&  Combination of hyperspectral imaging (reflection, mapping) and optical coherence tomography enabled by high brightness of the source employed & Aspects of combination discussed; hyperspectral imager combined with optical coherence tomography demonstrated; correlative imaging of polymers embedded in ceramics demonstrated &\makecell{ Zorin et al.\\ \cite{Zorin:20} 2020 }\\ 
\multicolumn{1}{>{\hspace{0pt}}m{0.02\columnwidth}}{} &  \multicolumn{1}{>{\hspace{0pt}}m{0.22\columnwidth}}{} & \multicolumn{1}{>{\hspace{0pt}}m{0.22\columnwidth}}{} & \multicolumn{1}{>{\hspace{0pt}}m{0.22\columnwidth}}{} & \multicolumn{1}{>{\hspace{0pt}}m{0.13\columnwidth}}{} \\
\bottomrule
\end{longtable}
\end{footnotesize}

\begin{figure}[H]
\centering
 \begin{subfigure}[t]{.35\columnwidth}
 \includegraphics[width=\columnwidth]{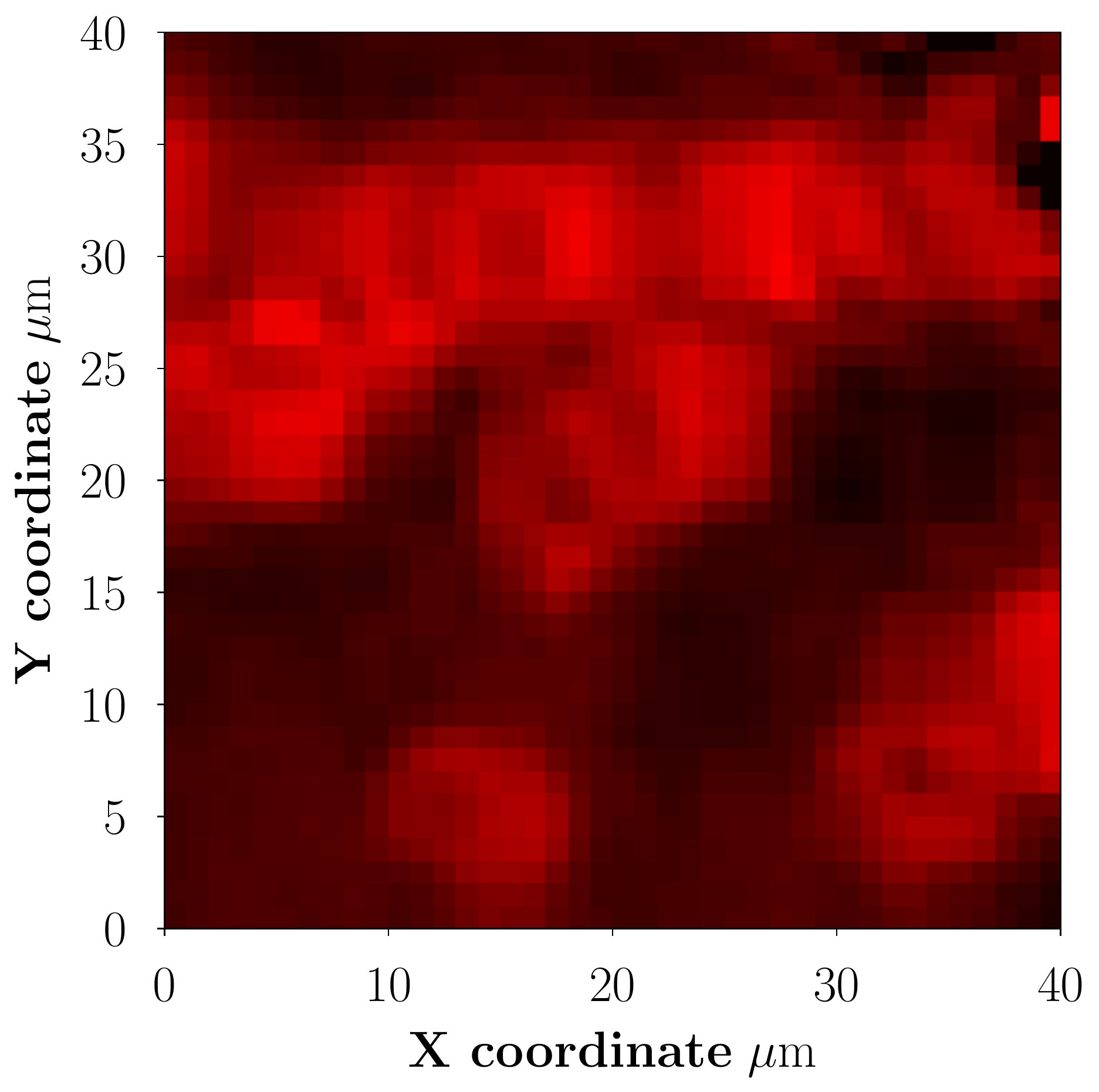} 
 \caption{Global intensity mid-IR microscopic image of red blood cells on substrate, obtained using mapping approach}\label{fig:blood-glob}
 \end{subfigure}
 \begin{subfigure}[t]{.35\columnwidth}
  \includegraphics[width=\columnwidth]{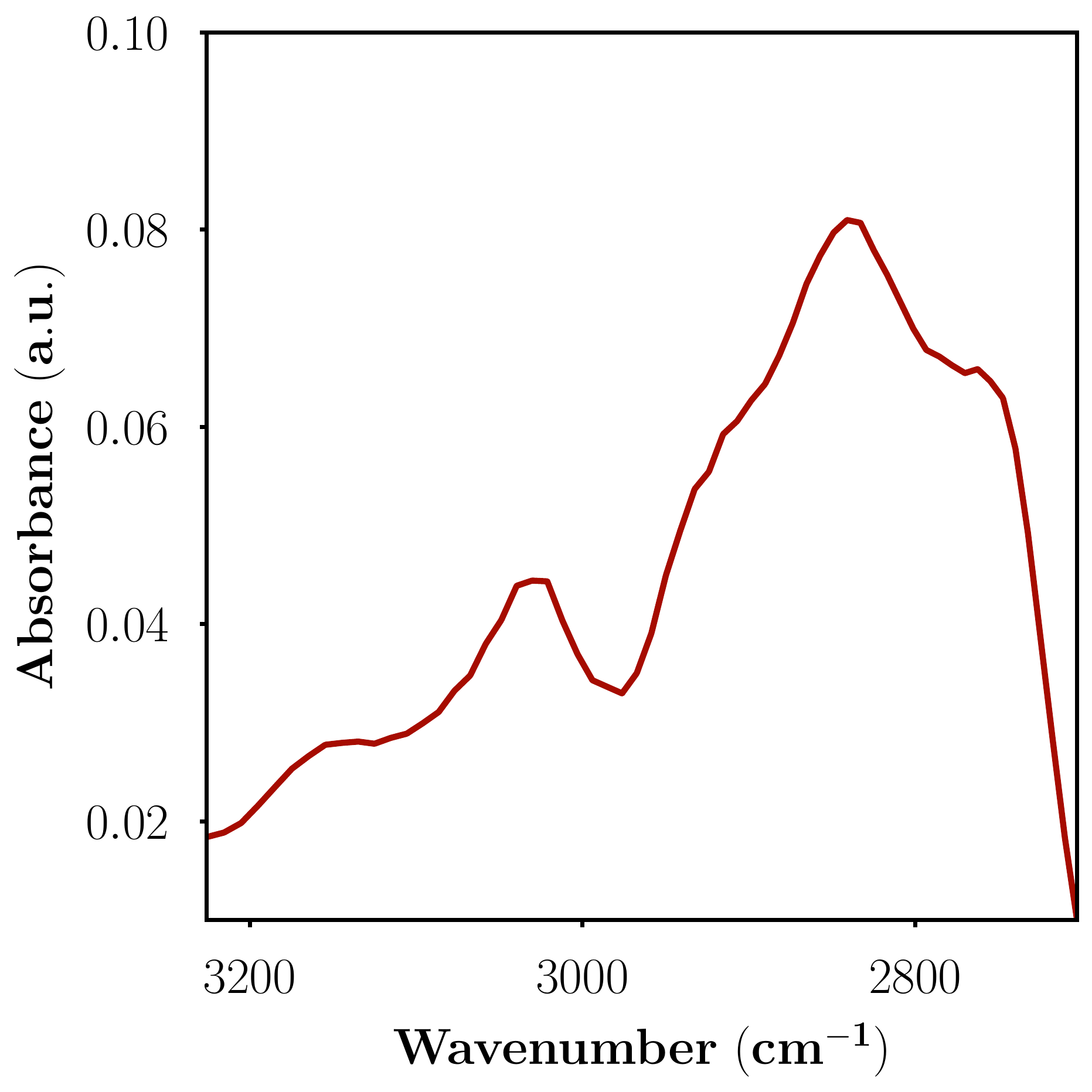}
  \caption{Absorption spectrum of a single red blood cell extracted from hyperspectral cube}\label{fig:bloodspectrum}
 \end{subfigure}
  \begin{subfigure}[t]{.65\columnwidth}
  \includegraphics[width=\columnwidth]{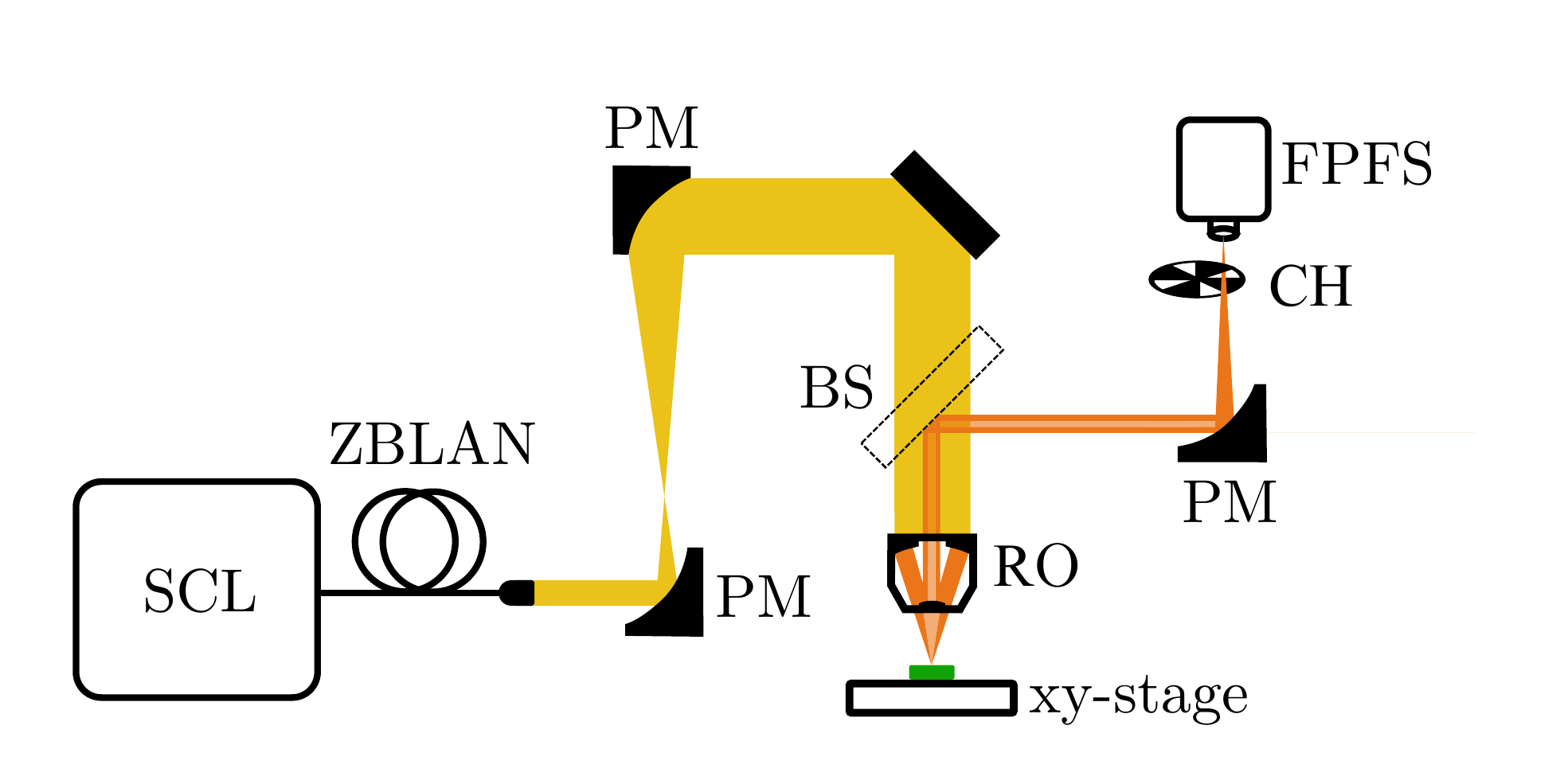}
  \caption{Scheme of the optical setup: SCL - supercontinuum Laser, PM - parabolic mirror, BS - beam splitter, RO - reflective objective, CH - chopper, FPFS - Fabry-P\'erot tunable filter spectrometer.}
 \end{subfigure}
  \caption{Diffraction-limited supercontinuum-based reflection microspectroscopy, reproduced from~\cite{Kilgus:18} with permission from OSA.}\label{fig:jakobs_setup}
 \begin{tikzpicture}[overlay,thick]
 \end{tikzpicture}
 \label{fig:jakob_blood}
\end{figure}

Notably, most of the applications listed in Table~\ref{tab:table2} distinguish and emphasize the particular importance of the high spectral brightness and beam quality (two sides of the same coin) of modern supercontinuum laser sources for imaging applications. In the publication of Kilgus et al.\cite{Kilgus:18}, mid-IR supercontinuum sources were demonstrated as a solution for eliminating the trade-off\textemdash specific for standard FTIR spectral microscopes\textemdash between spatial (resolution) and spectral (signal-to-noise ratio) performances. The practical aspects and signal-to-noise gains enabled by this new source type are discussed. Figure~\ref{fig:jakobs_setup} reproduces a layout of the all-mirror (mirror optics was employed to avoid chromatism) hyperspectral mid-IR microscope based on a supercontinuum source. The capabilities of the system were demonstrated by diffraction-limited imaging of erythrocytes. A corresponding image is depicted in  Fig.~\ref{fig:blood-glob}. Each pixel of the system's output contains a complete spectral information [see Fig.~\ref{fig:bloodspectrum}] that is accessed by a spectrometer (a Fabry-P\'erot tunable filter spectrometer operating from 3.1~\textmu m to 4.4~\textmu m). The reported spatial resolution of the system is around 4.5~\textmu m.

\begin{footnotesize}
\begin{longtable}{>{\hspace{0pt}}m{0.02\columnwidth}>{\raggedright\hspace{0pt}}m{0.22\columnwidth}>{\raggedright\hspace{0pt}}m{0.22\columnwidth}>{\raggedright\hspace{0pt}}m{0.22\columnwidth}>{\hspace{0pt}}m{0.13\columnwidth}}
\caption{Overview of supercontinuum based mid-IR stand-off/remote spectroscopy.\label{tab:table3}}\\ 
\toprule
\multicolumn{1}{>{\hspace{0pt}}m{0.02\columnwidth}}{\texttt{\#}} & \multicolumn{1}{>{\centering\hspace{0pt}}m{0.22\columnwidth}}{Type of supercontinuum} & \multicolumn{1}{>{\centering\hspace{0pt}}m{0.22\columnwidth}}{Application description} & \multicolumn{1}{>{\centering\hspace{0pt}}m{0.22\columnwidth}}{Details} & \multicolumn{1}{>{\centering\hspace{0pt}}m{0.13\columnwidth}}{Ref. Year} \endfirsthead 
\midrule
1 & Step-index ZBLAN fiber based, custom;  repetition rate 2~MHz (modulated at 500~Hz to reduce the thermal load), 3.9~W average power;   (13330-2325~cm\textsuperscript{-1})  & Stand-off diffuse reflection spectroscopy of explosives (TNT, RDX, PETN), fertilizers (ammonium nitrate, urea) and paints (automotive and military grade) & Stand-off system based on monochromator (InSb detector, lock-in demodu- lation) demonstrated; measurements (high quality spectra obtained and verified) at the distance of 5~m achieved, performances at distances up to 150~m predicted; importance of spectral extension beyond 6~\textmu m emphasized  &\makecell{ Kumar et al.\\  \cite{Kumar:12} 2012} \\ 
\midrule
2 & Step-index ZBLAN fiber based, commercial;  repetition rate 2.5~MHz, 490~mW average power;  (8330-2175~cm\textsuperscript{-1}) & Sensitive stand-off analysis of various samples (ammonium nitrate, transparent acrylic paint, black alkyd paint, 2-propanol) employing a low-cost MEMS Fabry-P\'erot spectrometer[see Fig.~\ref{fig:standoff-jakob-setup}] & Importance of filling the spectral gap of QCLs highlighted; robust design, verified stand-off spectroscopy and real-time capabilities [for monitoring the evapo- ration of 2-propanol, see Fig.~\ref{fig:standoff-jakob-spectrum}] at 5~m distances demonstrated; noise characterized &\makecell{ Kilgus et al.\\  \cite{doi:10.1177/0003702817746696} 2018} \\ 
\multicolumn{1}{>{\hspace{0pt}}m{0.02\columnwidth}}{} &  \multicolumn{1}{>{\hspace{0pt}}m{0.22\columnwidth}}{} & \multicolumn{1}{>{\hspace{0pt}}m{0.22\columnwidth}}{} & \multicolumn{1}{>{\hspace{0pt}}m{0.22\columnwidth}}{} & \multicolumn{1}{>{\hspace{0pt}}m{0.13\columnwidth}}{} \\
\bottomrule
\end{longtable}
\end{footnotesize}

\begin{figure}[ht]
\centering
 \begin{subfigure}[a]{.49\columnwidth}
 \includegraphics[width=\columnwidth]{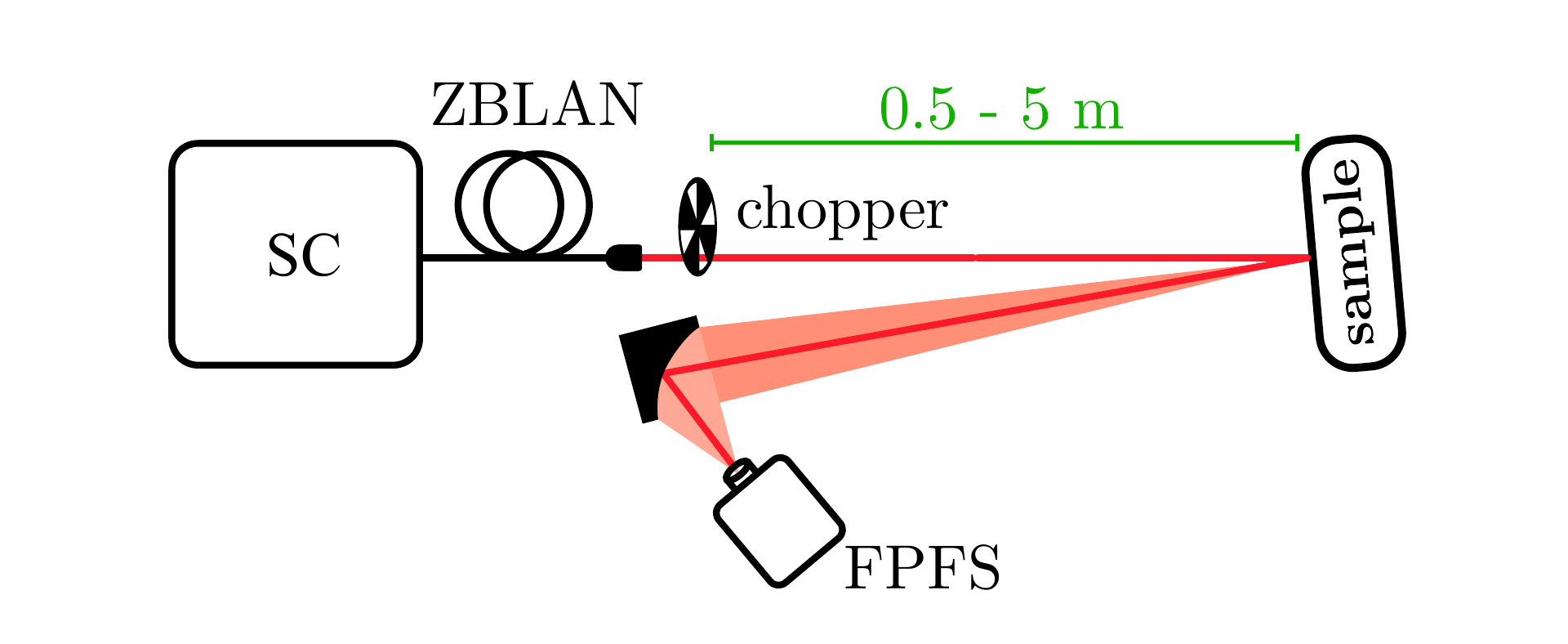} 
 \caption{Scheme of the supercontinuum-based stand-off detection system, SC - supercontinuum, FPFS -  MEMS Fabry-P\'erot spectrometer}\label{fig:standoff-jakob-setup}
 \end{subfigure}
 \begin{subfigure}[a]{.45\columnwidth}
  \includegraphics[width=\columnwidth]{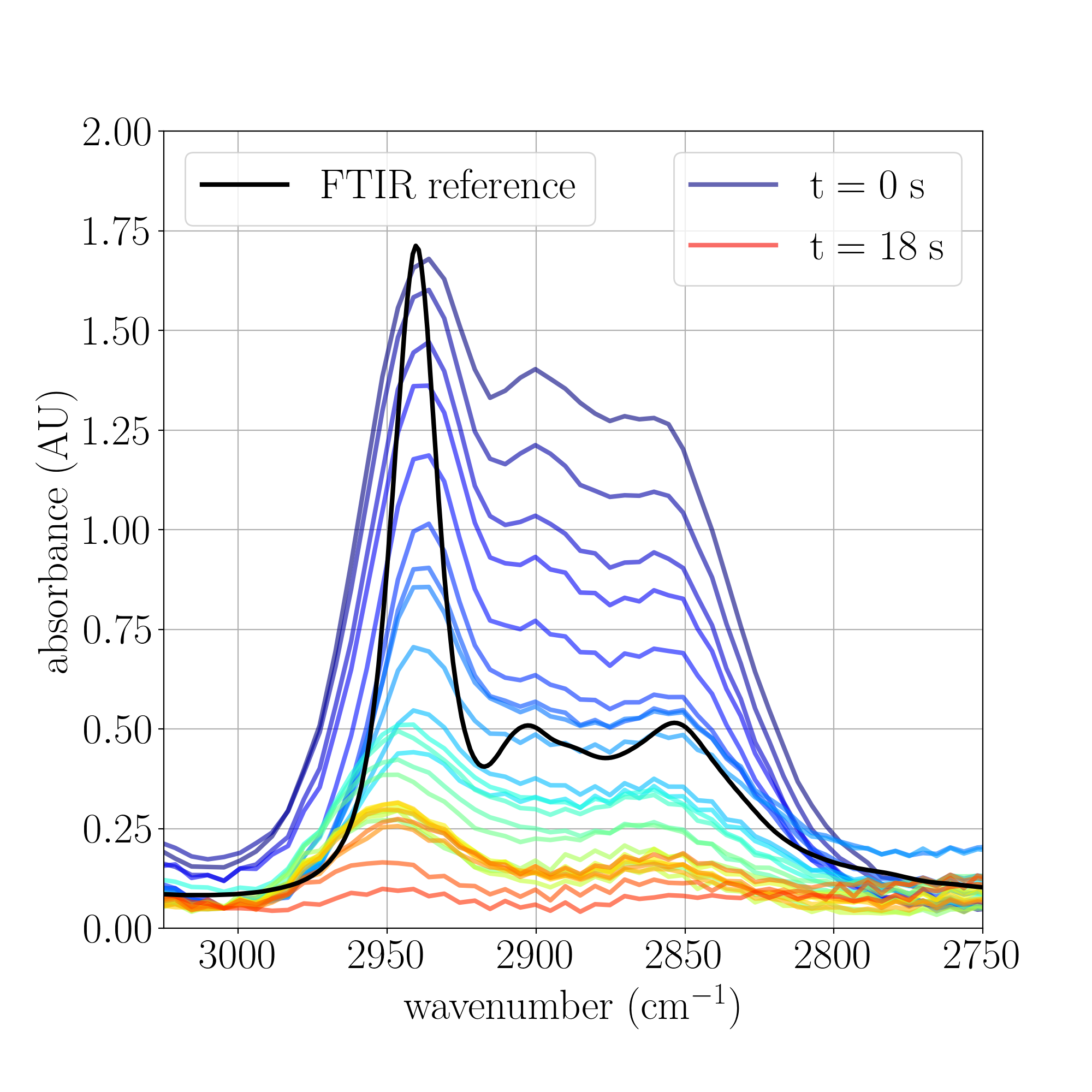}
  \caption{Time-resolved monitoring of the evaporation process of 2-propanol at 5~m distance (18~sec measurement)}\label{fig:standoff-jakob-spectrum}
 \end{subfigure}
  \caption{Supercontinuum-based mid-IR stand-off spectroscopy\cite{doi:10.1177/0003702817746696}, reproduced with permission from SAGE Publishing.}
 \begin{tikzpicture}[overlay,thick]
 \end{tikzpicture}
 \label{fig:standoff-jakob}
\end{figure}

The majority of the publications on the mid-IR supercontinuum generation imply potential advances for stand-off detection and remote spectroscopic sensing. Nevertheless, up-to-date only two reports (see Table~\ref{tab:table3}) were presented\cite{Kumar:12,doi:10.1177/0003702817746696}; both of them benefit from and illuminate the practical significance of the directionality, brightness, and spectral coverage of supercontinuum emission. Figure~\ref{fig:standoff-jakob} illustrates the capabilities of supercontinuum-based remote mid-IR spectroscopy exemplified for the latest publication (the optical scheme and time-resolved stand-off measurements of 2-propanol evaporation are displayed).

\newpage In order to provide a full snapshot on the current state-of-the-art, it is worth to include some near-IR realizations, which are relevant as important milestones in the development of IR supercontinuum sources. These works allow us to understand the state of supercontinuum-based IR spectroscopy in general and to extrapolate possible performance of similar mid-IR implementations that have not been yet reported in this range but expected. We summarise the most remarkable reports in Table~\ref{tab:tableX}.

\begin{footnotesize}
\begin{longtable}{>{\hspace{0pt}}m{0.02\columnwidth}>{\raggedright\hspace{0pt}}m{0.22\columnwidth}>{\raggedright\hspace{0pt}}m{0.22\columnwidth}>{\raggedright\hspace{0pt}}m{0.22\columnwidth}>{\hspace{0pt}}m{0.13\columnwidth}}
\caption{Overview of featured near-IR spectroscopic applications of supercontinuum sources.\label{tab:tableX}}\\ 
\toprule
\multicolumn{1}{>{\hspace{0pt}}m{0.02\columnwidth}}{\texttt{\#}} & \multicolumn{1}{>{\centering\hspace{0pt}}m{0.22\columnwidth}}{Type of supercontinuum} & \multicolumn{1}{>{\centering\hspace{0pt}}m{0.22\columnwidth}}{Application description} & \multicolumn{1}{>{\centering\hspace{0pt}}m{0.22\columnwidth}}{Details} & \multicolumn{1}{>{\centering\hspace{0pt}}m{0.13\columnwidth}}{Ref. Year} \endfirsthead 
\midrule
1 & Soft-glass photonic-crystal fiber based; custom, 142~MHz repetition rate, 25~mW average power; (14815-4545~cm\textsuperscript{-1})  & High resolution Fourier-transform absorption infrared spectroscopy (ammonia and acetylene overtone bands)& Performances of supercontinua for established techniques of IR spectroscopy as replacement of thermal sources demonstrated (high speed, high sensitivity achieved); significant impacts of possible extension towards mid-IR region stressed & \makecell{ Mandon et al.\\ \cite{Mandon:08} 2008} \\ 
\midrule
2 & Silica fiber based; 80~MHz repetition rate, 6~W average power; (22220-4000~cm\textsuperscript{-1})  & Near-IR Fourier transform spectroscopy (methane and methyl salicylate) using multipass absorption gas cell (18~m); in-deep noise characterization provided & High temporal stability of supercontinuum sources for time scales relevant to FTIR demonstrated and discussed; noise charac- terised in comparison to the tungsten lamp; drawbacks and potentials for mid-IR FTIR spectro- scopy distinguished & \makecell{ Michaels et al.\\ \cite{Michaels:09} 2009 }\\ 
\midrule
\newpage
\texttt{\#} &\centering Type of supercontinuum& \centering Application description&\centering Details&{\hspace{9pt} Ref. Year} \\ 
\midrule
3 & Standard single-mode silica fiber based (several amplification stages), custom; up to 20 MHz repetition rate, around 5~W average power; (6660-4255~cm\textsuperscript{-1}) & Reflectance spectroscopy of various samples (at 2~m distance); for sensing, beam delivered at distance of 1.6~km & Beam quality charac- terised at 1.6~km distance (Gaussian, symmetric, full angle beam divergence 0.49~mrad, M\textsuperscript{2}=1.26); active spectral reflection measurements of various samples demonstrated and verified; potentials as well as limitations (i.e. turbulence) discussed  &\makecell{ Alexander et al.\\  \cite{Alexander:13} 2013 }\\
\midrule
4 & Photonic-crystal fiber based, custom; up to 40~MHz repetition rate (10~MHz default), 1~W average power; high stability and repeatability stated; (10525-5880~cm\textsuperscript{-1}) & Supercontinuum laser absorption spectroscopy of several simple hydrocarbon species (methane, acetylene, ethylene, propane) at various concentrations and pressure conditions & Accuracy and feasibility of supercontinuum-based diagnostic strategy studied and verified; applicability of the technique demonstrated; potentials for steady-state
pyrolysis and gasification applications highlighted &\makecell{ Yoo et al.\\  \cite{doi:10.1177/0003702816641563} 2016}\\ 
\midrule
5 & Thulium-doped fiber based, commercial; 35~kHz repetition rate, 120 mW average power; (4760-3845~cm\textsuperscript{-1})  & Transmission spectroscopy of barley seeds using a supercontinuum laser & Prediction of mixed- linkage beta-glucan content in whole barley seeds with high accuracy demonstrated &\makecell{ Ringsted et al.\\ \cite{RINGSTED2017101} 2017  }\\ 
\midrule
6 & Standard single-mode silica fiber based, custom; 30 kHz repetition rate, pulse energy density ~25~nJ/nm (in 6060-5405~cm~\textsuperscript{-1} range); (6450-5265~cm\textsuperscript{-1}) & Spectroscopic photoacoustic imaging of lipids in the first overtone region using high pulse energy supercontinuum (based on high pulse energy ns pumping) & Low-cost high pulse energy supercontinuum developed and char- acterised; spectroscopic photoacoustic qualitative discrimination and cross-sectional scanning of lipids in the first overtone transition band of C-H bonds demonstrated &\makecell{ Dasa et al.\\  \cite{Dasa:18} 2018 }\\ 
\midrule
7 &  Standard single-mode silica fiber based, commercial; 250~kHz repetition rate, >1.3~W average power; (11110-3570~cm\textsuperscript{-1})\  & Photoacoustic spectroscopy for online multi-gas (carbon dioxide, methane, water vapor, hydrogen sulfide) sensing with high-speed data acquisition, resolution, and no interference from humidity & Suitability of super- continuum-based sensing for online monitoring for operational large scale biogas plants demonstrated &\makecell{Selvaraj et al.\\  \cite{doi:10.1021/acs.analchem.9b01513} 2019 }\\ 
\midrule
8 & Standard single-mode silica fiber based, custom; 280~kHz repetition rate, pumped by a 1~kW peak power system; (10000-6250~cm\textsuperscript{-1}) & Short-range supercontinuum lidar (up to 10~m) for spectroscopic temperature measurements in combustion units at plant environments (utilising absorption of water vapor) &  Gas absorption cross- sectional dependence on temperature exploited; advances over narrow-band lasers demonstrated; simultaneous 3D mapping of temperature and concentration for the combustion diagnosis proposed &\makecell{Saleh et al.\\  \cite{Saleh:19} 2019}\\
\midrule
\newpage
\texttt{\#} &\centering Type of supercontinuum& \centering Application description&\centering Details&{\hspace{9pt} Ref. Year} \\
\midrule
9 & Standard non-zero dispersion-shifted single-mode optical fiber based (several amplification stages), custom; 100 kHz repetition rate, pulse energy density ~18.3, 1830~mW average power \textmu J/nm; (6945-5350~cm\textsuperscript{-1}) & \textit{Ex vivo} (Adipose tissue) and \textit{in vivo} (\textit{Xenopus laevis} tadpoles, embryos) photoacoustic microscopy of lipids in the extended near-infrared using high pulse energy supercontinuum & High pulse energy supercontinuum developed for multi- spectral photoacoustic microscopy; high quality imaging over first overtone transition of C–H vibration bonds demonstrated &\makecell{ Dasa et al.\\  \cite{DASA2020100163} 2020  }\\ 
\midrule
10 & Standard single-mode optical fiber based, custom; 30-100 kHz repetition rate, around 20~mW average power (after gas cell filled with pure N\textsubscript{2}); (6755-5880~cm\textsuperscript{-1}) & Compact, low-cost all-fiber supercontinuum-based gas sensor for detection of multiple industrial toxic gasses & System for multi-species gas detection (ammonia, methane) reliable for continuous monitoring with high selectivity and sensitivity (4~ppm) demonstrated; &\makecell{ Adamu et al.\\  \cite{9090856} 2020}\\ 
\multicolumn{1}{>{\hspace{0pt}}m{0.02\columnwidth}}{} &  \multicolumn{1}{>{\hspace{0pt}}m{0.22\columnwidth}}{} & \multicolumn{1}{>{\hspace{0pt}}m{0.22\columnwidth}}{} & \multicolumn{1}{>{\hspace{0pt}}m{0.22\columnwidth}}{} & \multicolumn{1}{>{\hspace{0pt}}m{0.13\columnwidth}}{} \\
\bottomrule
\end{longtable}
\end{footnotesize}

\begin{figure}[ht]
\centering
 \begin{subfigure}[a]{.49\columnwidth}
 \includegraphics[height=5cm]{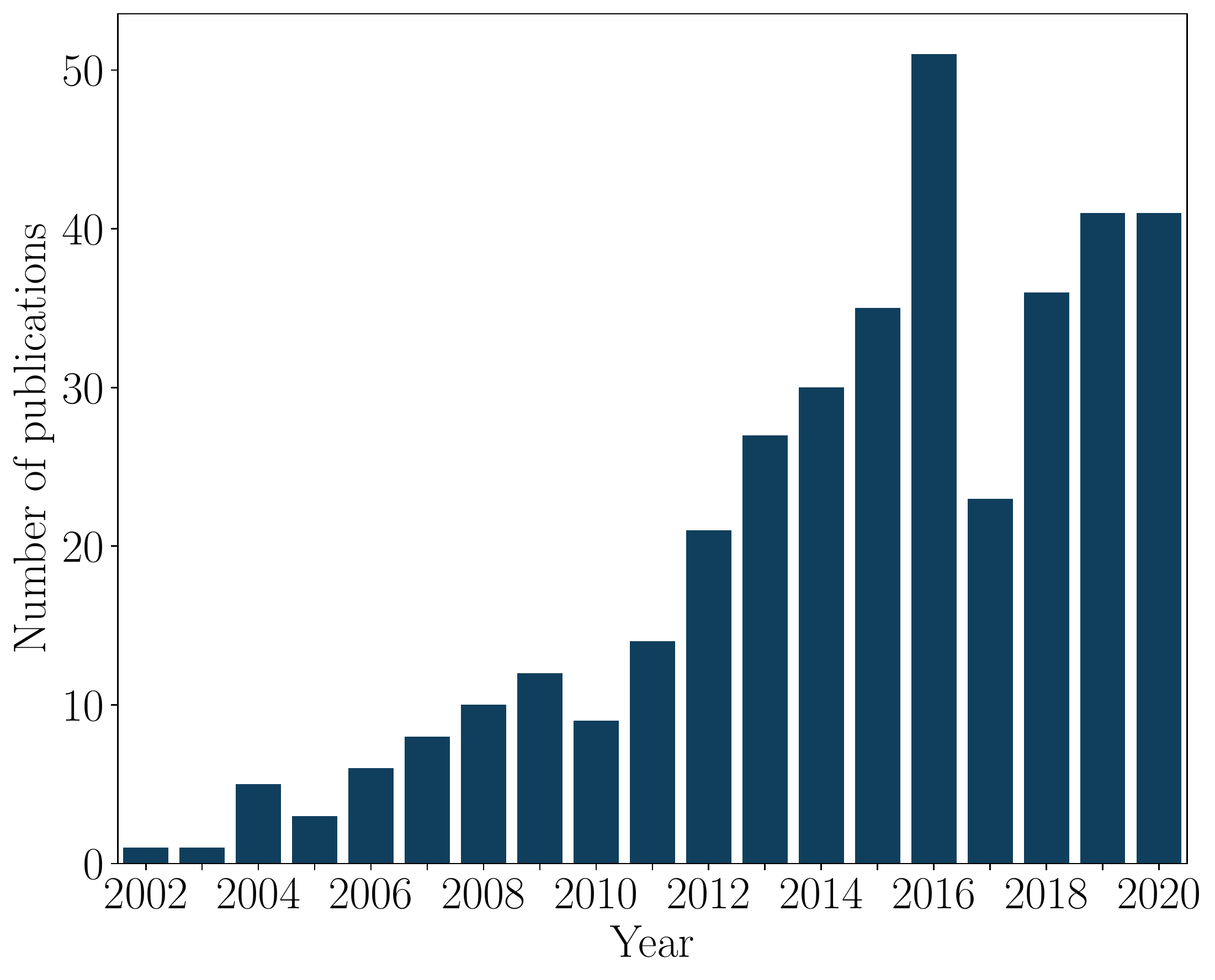} 
 \caption{}\label{fig:pbmed}
 \end{subfigure}
 \begin{subfigure}[a]{.49\columnwidth}
  \includegraphics[height=5cm]{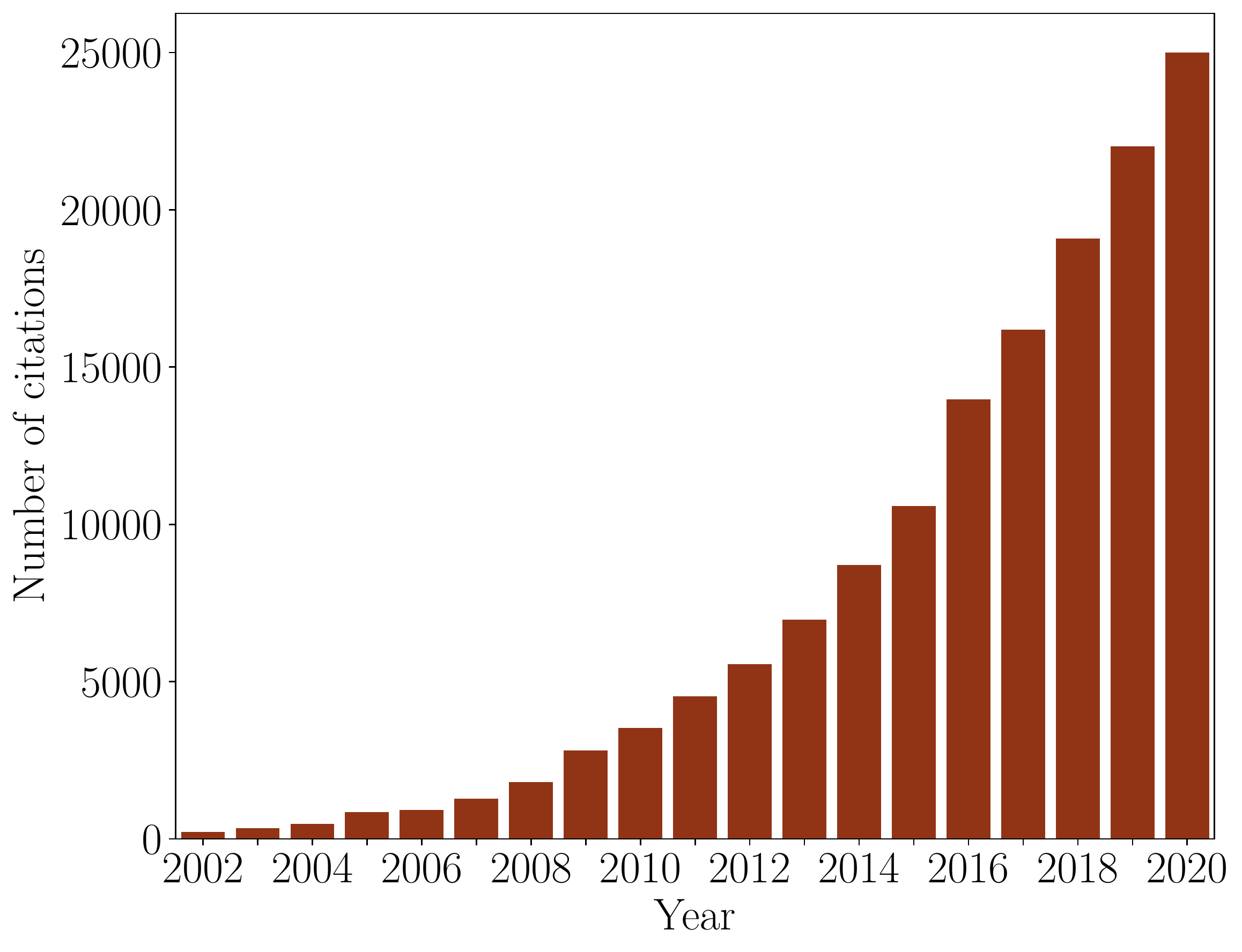}
  \caption{}\label{fig:dim}
 \end{subfigure}
  \caption{Current technology state and academic impact: (a) Number of scientific publications by year that contain keywords "infrared supercontinuum", according to the "PubMed" search (primarily biomedical literature); (b) Citations per year (the number of times that a publication has been cited by other publications) related to the search "mid-infrared supercontinuum", according to "Dimensions" research information database.}
 \begin{tikzpicture}[overlay,thick]
 \end{tikzpicture}
 \label{fig:pubnum}
\end{figure}

Analyzing the tables, we can note that the number of publications related to mid-IR spectroscopy is increasing annually, indicating a growing interest and development in this field. Figure~\ref{fig:pubnum} complements this analysis and shows the state of technology and academic impact as measured by the number of publications and citations. We believe that the current progress is primarily driven by developments and commercialization of novel high-performance supercontinuum laser sources, i.e. their extending availability. 


Nevertheless, the absolute figures show that despite the growing number of reports on supercontinuum generation (prevailing fraction of the scientific publications) and its advances in the mid-IR, the application of this technology is still emerging at the step of being noticed and adopted by a broader community of spectroscopists. At the moment, the research is still very much dominated by single groups, which might explain the unsteady development of published papers (e.g. the drop in 2017).
In addition, we can note that reported applications in the near-IR and shortwave mid-IR still prevail over long-wavelength realizations. We attribute that to the lag in commercialization of ultra-broadband sources, although, systems covering the entire mid-IR fingerprint region were reported and successfully demonstrated. However, the situation is expected to change as first commercial systems with the extended coverage appear\cite{noauthor_norblis_nodate,noauthor_leukos_nodate}.
Figure~\ref{fig:pubnum} was compiled using open data sources. The data on the number of publications (primarily biomedicine related literature) were obtained from the PubMed database (the search keywords are "infrared supercontinuum"); the data on the number of citations (for the keywords "mid-infrared supercontinuum") were obtained using "Dimensions" research information database\cite{dimensions}.

\section{Conclusions and outlook}

This review has set a focus on various practical aspects and prospects of supercontinuum technology that are relevant for mid-IR spectroscopy.
Just as progress on the supercontinuum generation in the 1970s was driven by the interests of Raman spectroscopy, we find that current progress is accelerated primarily by interests of IR spectroscopy, although only a limited number of groups had hands on this novel type of mid-IR laser source at the time of writing.
In this contribution, we aimed to introduce and demonstrate specific characteristics of supercontinuum emitters to the broad spectroscopic community. For this purpose, we have analyzed, quantified and illustrated typical emission properties of several commercial mid-IR supercontinuum sources. Thus, spectral brightness, coverage, laser emission properties (in particular the M\textsuperscript{2} beam quality), and stability (long- and short-term) have been investigated. In particular cases, we have provided a comparison of the supercontinuum technology to standard, well-established and advanced mid-IR spectroscopic equipment to highlight the emerging capabilities.
In a detailed overview, we have summarized and assessed the current state-of-the-art with a focus on highly promising reports and achievements in the field of mid-IR spectroscopy. 
The review shows that the technology of mid-IR supercontinuum generation offers novel, previously unavailable features and is sufficiently mature to enter the applied field. The technology can compete in well-known application scenarios, but it can also create new analytical methods. Thus, mid-IR supercontinuum laser sources are candidates to at least fill the gap between quantum cascade lasers (QCL) and standard thermal emitters or even potentially overcome the latter in particular cases. The main advantages are the ultra-broadband spectral coverage, which also covers bands not available with QCL technology, and superior brightness levels at a reasonable price per spectral band. In addition, the surveyed work also shows the need for both further fundamental and applied research in this field. Further extensions of the spectral bandwidth, commercialization of the sources covering the entire fingerprint region, improvements of noise performance through tailoring the supercontinuum generation dynamics are of particular importance. These steps will significantly stimulate the application field.



\section*{Acknowledgments}
The authors thank Guillaume Huss from Leukos for the fruitful and detailed discussions and providing the power spectral densities of various mid-IR supercontinuum sources used to calculate the spectral brightness levels; Roozbeh Shokri and Reza Salem from Thorlabs for discussions and lending the InF\textsubscript{3} fiber based supercontinuum source; NKT Photonics and particularly Patrick Bowen for technical conversations and providing a low-cost mid-IR ZBLAN based source; and Novae, in particular Marc Castaing, Nicolas Ducros and Kirill Zaytsev, for discussions and providing the mid-IR supercontinuum generator (Novae Coverage) for our experiments and characterization. In addition, Ivan Zorin acknowledges former SUPUVIR partners (an abbreviation for "SUPercontinuum broadband light sources covering UV to IR applications") for establishing a firm scientific network in developments and applications of supercontinuum sources; Ivan is grateful to be part of the community that generated many of the works reviewed in this contribution. This work was co-financed by research subsidies granted by the government of Upper Austria.

\section*{Funding}
{\"O}sterreichische Forschungsf{\"o}rderungsgesellschaft (FFG) (877481, 874787, 856896); State of Upper Austria (Wi-2020-700476/3); Horizon 2020 Framework Programme (722380).

\section*{Disclosures}
The authors declare no conflicts of interest.

\section*{Data availability}
Data underlying the results presented in this paper are not publicly available at this time but may
be obtained from the authors upon reasonable request.


\bibliography{sample}

\end{document}